\documentclass[journal]{IEEEtran}

\usepackage{cite}
\usepackage{amsmath,amssymb,amsfonts}
\usepackage{algorithm}
\usepackage{algpseudocode}
\usepackage{graphicx}

\usepackage{textcomp}
\usepackage{xcolor}
\usepackage{bm}
\usepackage{booktabs}
\usepackage{multirow}
\usepackage{multicol}
\usepackage{longtable}
\usepackage{array}

\usepackage{flushend}
\usepackage[caption=false,font=footnotesize]{subfig}
\usepackage{url}
\usepackage{tabularx}
\usepackage{float}
\usepackage{stfloats}
\usepackage{graphicx}
\usepackage{pifont}
\usepackage{placeins}
\newcommand{\cmark}{\ding{51}}
\newcommand{\xmark}{\ding{55}}

\usepackage{amsthm}

\begin{document}

\title{Hybrid PPO-Based Multi-Objective Optimization for UAV-mounted STAR-RIS with Energy Harvesting}

\author{
Anas Hussin,
Khaled M. Rabie,~\IEEEmembership{Senior Member,~IEEE},
Mahmoud M. Salim, Ali H. Muqaibel~\IEEEmembership{Senior Member,~IEEE}, and Louai A. Al-Awami
~\IEEEmembership{Senior Member,~IEEE}

\thanks{Anas Hussin, Khaled M. Rabie, and Louai A. Al-Awami are with the Computer Engineering Department, King Fahd University of Petroleum and Minerals (KFUPM), Dhahran 31261, Saudi Arabia. Mahmoud M. Salim and Ali H. Muqaibel are with the Electrical Engineering Department, KFUPM, Dhahran 31261, Saudi Arabia.}

\thanks{Khaled M. Rabie, Louai, Mahmoud M. Salim, and Ali H. Muqaibel are also with the Center for Communication Systems and Sensing, KFUPM, Dhahran 31261, Saudi Arabia.}

\thanks{Louai A. Al-Awami is also with the Interdisciplinary Research Center for Intelligent Secure Systems, KFUPM, Dhahran 31261, Saudi Arabia.}

\thanks{Corresponding authors: Anas Hussin and Khaled M. Rabie (e-mail: g202321390@kfupm.edu.sa; k.rabie@kfupm.edu.sa).}
%
%
%
%
}

\markboth{IEEE Transactions on XXXXX,~Vol.~XX, No.~XX, Month~Year}%
{First Author \MakeLowercase{\textit{et al.}}: Short Title of Paper}

\maketitle

\begin{abstract}
This paper investigates a multi-objective optimization framework for UAV-mounted reconfigurable intelligent surface (STAR-RIS)-assisted Sixth Generation (6G)-enabled Internet of Things (IoT) systems. The considered framework jointly supports communication enhancement and radio-frequency (RF) energy harvesting (EH) through adaptive STAR-RIS transmission and reflection control. A weighted multi-objective optimization problem is formulated to jointly optimize the base station (BS) beamforming vectors, STAR-RIS transmission and reflection coefficients, RF power-splitting factor, and UAV trajectory under quality-of-service (QoS), power, and mobility constraints. Due to the strong coupling among the communications, EH, and UAV mobility variables, the resulting problem is highly nonconvex and difficult to solve using conventional optimization methods. To address this challenge, a hybrid proximal policy optimization (H-PPO) framework is proposed. The developed hybrid actor-critic architecture combines a categorical policy for discrete UAV movement decisions with Beta-distributed policies for continuous communication-control variables. In addition, a two-stage reward mechanism is introduced to enforce QoS feasibility while balancing system sum rate, energy efficiency, and harvested energy. Simulation results demonstrate that the proposed H-PPO framework achieves faster convergence and more stable training behavior than proximal policy optimization (PPO) and twin delayed deep deterministic policy gradient (TD3) baselines. In particular, H-PPO achieves approximately 66\% and 60\% faster convergence than PPO and TD3, respectively, while maintaining more stable training behavior. Furthermore, the proposed framework effectively balances communication performance and EH under different QoS requirements, STAR-RIS sizes, transmit power levels, and user densities.
\end{abstract}


\section{Introduction}
The Internet of Things (IoT) is expected to be an essential pillar of the future Sixth Generation (6G) wireless systems, enabling massive connectivity among intelligent devices, autonomous systems, sensors, etc. As such networks evolve, more intelligent, flexible, and energy-efficient services are expected \cite{10361836}, \cite{11207621}. However, traditional terrestrial infrastructures suffer from limited coverage, reduced flexibility, and performance degradation in obstructed environments. This particularly takes place in dense urban and dynamic scenarios where severe path loss and blockages degrade quality of service (QoS). To address these challenges, reconfigurable intelligent surfaces (RIS) have emerged as a promising technology for next-generation wireless networks \cite{9110869}.

RIS consists of programmable metasurfaces composed of passive elements capable of dynamically adjusting the phase and amplitude of incident electromagnetic waves \cite{11215794}, \cite{10051712}. This capability enables the wireless propagation environment to be reconfigured, allowing signals to be directed toward intended users while mitigating interference. Consequently, RIS can significantly improve spectral and energy efficiency while preserving low hardware complexity and power consumption compared to conventional active relaying techniques \cite{10051712}, \cite{10704600}. However, conventional RIS architectures are limited to reflecting signals on a single side, which limits their spatial coverage and resource utilization.

To overcome this limitation, simultaneous transmitting and reflecting RIS (STAR-RIS) has been introduced, enabling each element to simultaneously transmit and reflect incident signals \cite{9964251}. This capability provides full-space coverage and improves the utilization of spatial resources, thereby enhancing overall network performance and user fairness \cite{10306287}. Despite these benefits, implementing STAR-RIS in fixed terrestrial setups continues to encounter difficulties, particularly in terms of adaptable coverage and rapidly changing environmental conditions.

The combination of unmanned aerial vehicles (UAVs) with RIS/STAR-RIS technologies provides a flexible and better solution to these challenges. UAV-assisted RIS systems enable dynamic three-dimensional deployment, facilitating the establishment of reliable line-of-sight (LoS) links and extending coverage in complex environments \cite{10539342}, \cite{11174947}. By adaptively adjusting their trajectories and surface configurations, UAV-mounted STAR-RIS platforms can reduce path loss and significantly enhance communication performance. Such systems are particularly suitable for applications such as emergency communications, dense urban scenarios, and rapidly changing network conditions \cite{10704600}.

However, enhanced flexibility and performance of UAV-assisted STAR-RIS systems are accompanied by higher energy consumption, thereby making energy sustainability a key issue. To address this issue, energy harvesting (EH) mechanisms have been incorporated into RIS architectures, enabling the surface elements to harvest a portion of the received radio-frequency (RF) energy while utilizing the remaining portion for communications \cite{10051712}, \cite{KHALIL2023109784}, \cite{10452297}. This introduces a fundamental trade-off between communication performance and harvested energy, as increasing EH reduces the power available for signal reception. Consequently, the resulting system design becomes highly coupled and non-convex, posing significant challenges for conventional optimization approaches.

To effectively manage these complex interactions, deep reinforcement learning (DRL) has proposed as a powerful framework for solving high-dimensional and dynamic optimization problems \cite{10704600}. DRL enables model-free learning, allowing an intelligent agent to adaptively optimize system performance through interaction with the environment. In UAV-assisted STAR-RIS networks, DRL-based algorithms such as deep deterministic policy gradient (DDPG), twin delayed deep deterministic policy gradient (TD3), and proximal policy optimization (PPO) have demonstrated strong capabilities in jointly optimizing UAV trajectory, transmit power, and surface configuration to enhance communication performance and energy efficiency \cite{10704600}, \cite{10306287}, \cite{CHAUDHARY2025700}, \cite{10187161}, \cite{10458888}. These approaches enable continuous adaptation to varying channel conditions and system requirements while balancing multiple performance objectives.

\subsection{Related Work}

\subsubsection{STAR-RIS-Assisted Communications}

STAR-RIS extends conventional RIS by enabling simultaneous transmission and reflection, thereby achieving full-space signal coverage and improved spatial resource utilization \cite{10306287}, \cite{10458888}, \cite{10187159}. In non-orthogonal multiple access systems, STAR-RIS has been used to improve energy efficiency through joint optimization of transmission power, user scheduling, and surface coefficients under QoS and fairness constraints \cite{9964251}. Coverage-capacity trade-offs have also been addressed using DRL-based frameworks that dynamically adapt STAR-RIS configurations \cite{10187159}. Moreover, hardware-aware designs have been explored using coupled phase-shift models to reflect practical STAR-RIS limitations \cite{9837935}. In vehicular networks, STAR-RIS has been applied to enhance signal coverage and mitigate blockage effects under dynamic channel conditions \cite{CHAUDHARY2025700}. These studies highlight the effectiveness of STAR-RIS in improving communication performance through flexible signal propagation control.  Nonetheless, the surface deployment in \cite{10306287,10458888,10187159,9964251,9837935,CHAUDHARY2025700} is either fixed or restricted to ground-vehicle mobility, and none of these works jointly account for RF energy harvesting at the surface, which limits their applicability to energy-constrained, dynamically deployable IoT scenarios.

\subsubsection{UAV-Assisted Wireless Systems}

UAVs have been widely studied as a flexible solution for enhancing wireless communication coverage and reliability. Their ability to dynamically adjust altitude and position enables the establishment of strong LoS links, making them suitable for complex and obstructed environments \cite{10051712}, \cite{10704600}, \cite{10539342}, \cite{10458888}. UAV-assisted systems have been applied in various contexts, including mobile edge computing, vehicular networks, and wireless sensor networks. In these systems, UAV trajectory optimization has been used to improve communication performance, reduce latency, and enhance energy efficiency \cite{10704600}, \cite{10802901}. These studies demonstrate the effectiveness of UAV mobility as a key enabler for next-generation wireless networks. However, the majority of these UAV-assisted designs \cite{10051712,10704600,10539342,10802901} rely on conventional single-sided RIS or omit the reconfigurable surface altogether, thereby underutilizing the full-space coverage that STAR-RIS can provide once mounted aerially. 

\subsubsection{EH in Wireless Systems}

EH plays a critical role in improving the sustainability of wireless systems, particularly in energy-constrained scenarios. In simultaneous wireless information and power transfer systems, EH enables receivers to extract energy from RF signals while maintaining communication functionality \cite{10051712}, \cite{10452297}. Advanced EH models have been proposed to capture practical circuit behavior. For instance, the quasi-concave S-type model in \cite{11176824} reflects both growth and saturation characteristics of EH circuits, enabling more accurate optimization of beamforming and power-splitting variables. Time-switching and power-splitting protocols have also been developed to dynamically allocate resources between EH and information transmission \cite{11214186}. In integrated sensing and communication systems, EH has been used to support both communication and sensing functionalities \cite{10891707}. These works highlight the importance of adaptive EH mechanisms in improving system reliability and efficiency. Yet, these EH designs \cite{11176824,11214186,10891707} are developed for terrestrial or single-function surfaces, and do not consider the additional coupling introduced when EH, STAR-RIS operation, and UAV mobility are optimized jointly.

\subsubsection{DRL-Based Optimization for RIS Systems}

DRL has emerged as an effective tool for solving high-dimensional and non-convex optimization problems in wireless communications \cite{9110869}, \cite{9964251}, \cite{10458888}. In RIS-assisted systems, DRL has been used to optimize phase shifts and beamforming strategies, achieving faster convergence compared to conventional optimization techniques \cite{9110869}. In UAV-assisted scenarios, DRL has been applied to adapt system parameters under dynamic channel conditions and energy constraints \cite{10793392}. Various DRL algorithms, including DDPG, TD3, and PPO, have been used to optimize communication performance in RIS and STAR-RIS systems \cite{10306287}, \cite{10187161}, \cite{9741332}. Recent works have also explored enhancements such as entropy regularization, action bounding, and multi-agent learning to improve stability and scalability \cite{11174947}, \cite{9645164}, \cite{10439000}. However, most existing approaches treat the control problem as fully continuous, without explicitly addressing structured or hybrid action-space formulations. This limitation is significant because several of the control variables in a UAV-mounted STAR-RIS system, most notably UAV movement, are naturally discrete or categorical, so purely continuous DRL formulations \cite{9110869,10793392,10306287,10187161,9741332,11174947,9645164,10439000} do not exploit the underlying structure of the action space and may consequently learn less efficiently.

\subsection{Motivations and Contributions}

\begin{table*}[!t]
\centering
\caption{Comparison of Existing Works}
\label{tab:lit_comparison}
\footnotesize
\setlength{\tabcolsep}{3.5pt}
\renewcommand{\arraystretch}{1.15}

\begin{tabular}{@{}p{1.5cm}p{3.4cm}ccccp{2.5cm}@{}}
\toprule
\textbf{Reference} & \textbf{Main Focus} & \textbf{STAR-RIS} & \textbf{EH} & \textbf{Mobility} & \textbf{UAV-mounted-STAR-RIS} & \textbf{Method} \\
\midrule

\cite{10051712} & EH-aware UAV-RIS optimization & \xmark & \cmark & \cmark & \xmark & Robust DRL (SD3) \\
\cite{10458888} & Aerial STAR-RIS for MEC & \cmark & \xmark & \cmark & \cmark & PPO \\
\cite{9964251} & EE optimization in NOMA STAR-RIS & \cmark & \xmark & \xmark & \xmark & DDPG \\
\cite{10704600} & EE RIS-assisted UAV networking & \xmark & \cmark & \cmark & \xmark & TD3 \\
\cite{10306287} & STAR-RIS-assisted V2X resource allocation & \cmark & \xmark & \xmark & \xmark & DDQN + Optimization \\
\cite{CHAUDHARY2025700} & Sum-rate in vehicular STAR-RIS & \cmark & \xmark & \cmark & \xmark & PPO \\
\cite{10187159} & Coverage-capacity optimization & \cmark & \xmark & \xmark & \xmark & MO-PPO \\
\cite{10802901} & Multi-agent EE optimization & \cmark & \xmark & \cmark & \xmark & Multi-agent DRL \\
\cite{11174947} & UAV-RIS with EH and jitter & \xmark & \cmark & \cmark & \xmark & SSD3 \\
\midrule
\textbf{This work} & \textbf{Joint multi-objective STAR-RIS-UAV-EH optimization} & \cmark & \cmark & \cmark & \cmark & \textbf{Hybrid PPO} \\
\bottomrule
\end{tabular}
\end{table*}

As summarized in Table~\ref{tab:lit_comparison}, existing studies primarily focus on optimizing subsets of system components rather than addressing a fully integrated framework. While some works incorporate EH and UAV mobility \cite{10051712}, \cite{10704600}, they do not consider STAR-RIS-enabled communications. Conversely, several studies focus on STAR-RIS optimization \cite{9964251}, \cite{10306287}, \cite{10187159}, but assume static deployments without UAV mobility or EH. Although UAV-assisted STAR-RIS systems have been explored \cite{10458888}, \cite{CHAUDHARY2025700}, \cite{10802901}, these works typically address single-objective optimization or partially coupled designs. Moreover, existing DRL-based approaches predominantly treat the action space as fully continuous, without explicitly leveraging structured or hybrid control designs to improve learning efficiency.

In contrast, this work develops a unified framework that jointly considers STAR-RIS configuration, UAV mobility, and EH under a multi-objective formulation. Furthermore, a hybrid action-space design is introduced within a PPO-based DRL framework to improve optimization performance and training stability in high-dimensional control problems. The key contributions of this research are highlighted below:

\begin{itemize}

\item {A comprehensive system model is developed for a UAV-mounted STAR-RIS-enabled wireless network supporting both transmission-side and reflection-side users, while incorporating RF-EH. The model captures the joint effects of BS beamforming, STAR-RIS transmission and reflection coefficients, RF power splitting, and UAV mobility under practical constraints.}

\item {The formulation of the multi-objective optimization problem involves finding a solution that maximizes both the system sum rate, energy efficiency, and the EH while satisfying QoS, transmit power, STAR-RIS hardware limitations, and mobility of UAV constraints.}

\item {A novel approach for controlling UAVs using a hybrid PPO method is presented, where the action space is explicitly designed to be hybrid, comprising both discrete and continuous parameters. In particular, the discrete actions are modeled using a categorical policy, whereas continuous actions are optimized using Beta-distributed policies.}

\item {A two-stage reward mechanism is introduced to first enforce QoS feasibility and subsequently balance multiple performance objectives through normalized reward components, improving training stability and policy effectiveness.} 
\item Simulation results validate the effectiveness of the proposed H-PPO framework in jointly optimizing communication performance and EH in UAV-mounted STAR-RIS systems, while consistently outperforming conventional PPO and TD3 benchmarks in terms of convergence speed, training stability, and overall multi-objective performance.

\end{itemize}




\section{System Model}
\label{sec:system_model}

\begin{figure}[!t]
\centering
\includegraphics[width=0.9\linewidth]{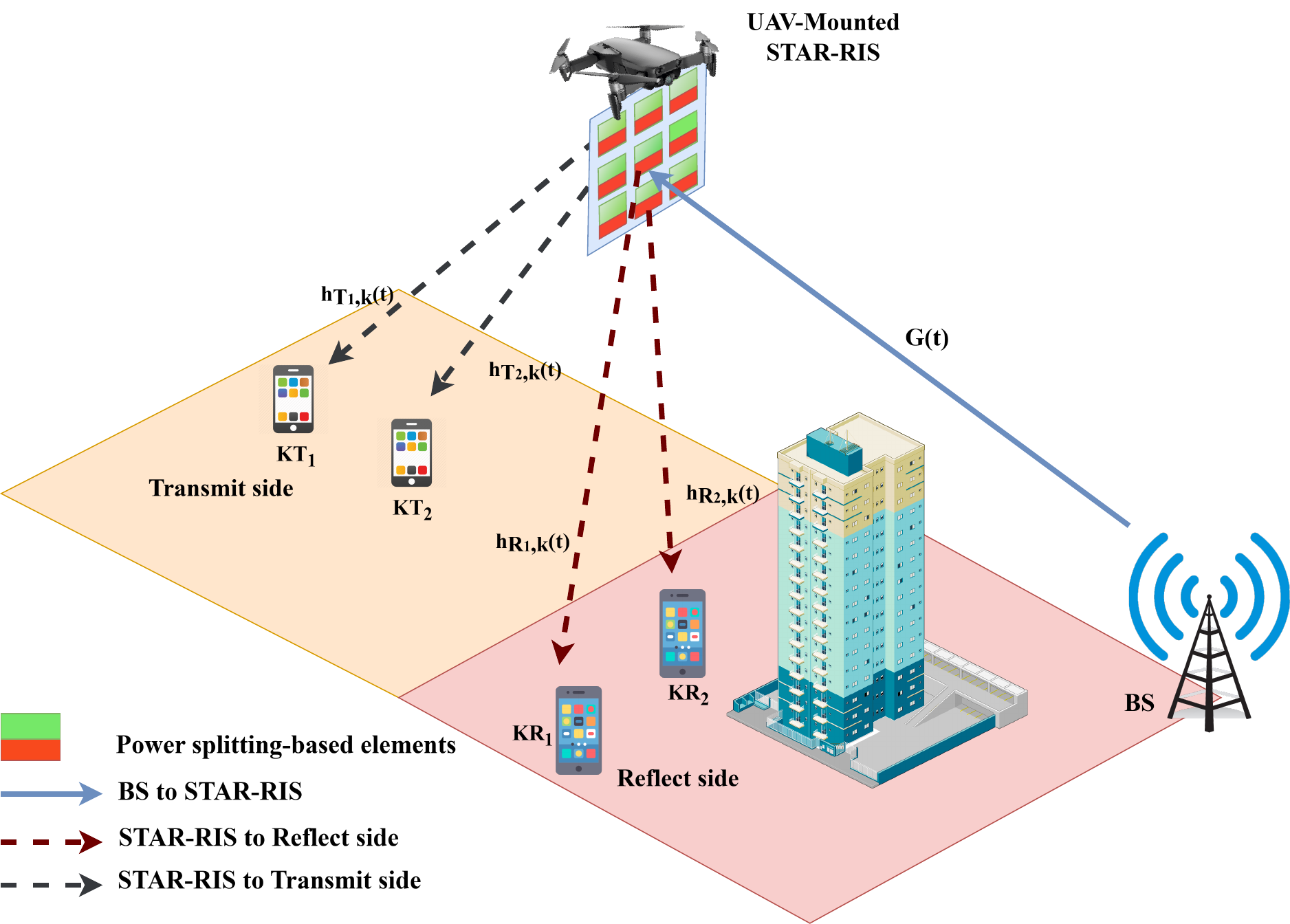}
\caption{UAV-mounted STAR-RIS assisted communication system with transmission-side and reflection-side users.}
\label{fig:system}
\end{figure}

As illustrated in Fig.~\ref{fig:system}, we consider a UAV-mounted STAR-RIS-assisted system, where a multi-antenna BS serves multiple user terminals (UTs) via an aerial STAR-RIS. The STAR-RIS consists of $N$ passive elements and is mounted on a UAV to enable flexible deployment. Due to blockage (e.g., buildings), the direct BS-UT links are assumed to be unavailable, and communication is established through cascaded BS-STAR-RIS-UT links. The STAR-RIS serves two groups of users located on its transmission and reflection sides, denoted by $\mathcal{K}_{\mathrm{T}}$ and $\mathcal{K}_{\mathrm{R}}$, respectively. The overall user set is $\mathcal{K}=\mathcal{K}_{\mathrm{T}}\cup \mathcal{K}_{\mathrm{R}}$, with cardinality $K=K_{\mathrm{T}}+K_{\mathrm{R}}$, where $K_{\mathrm{T}}=|\mathcal{K}_{\mathrm{T}}|$ and $K_{\mathrm{R}}=|\mathcal{K}_{\mathrm{R}}|$. The BS is equipped with $M$ antennas and located at $\mathbf{q}_{\mathrm{B}}=(x_{\mathrm{B}},y_{\mathrm{B}},H_{\mathrm{B}})$. The UAV flies at a fixed altitude $H_{\mathrm{U}}$, and its position at time slot $t$ is $\mathbf{q}_{\mathrm{U}}(t)=(x_{\mathrm{U}}(t),y_{\mathrm{U}}(t),H_{\mathrm{U}})$. The positions of transmission-side and reflection-side users are denoted by $\mathbf{q}_{k}^{\mathrm{T}}(t)=(x_{k}^{\mathrm{T}}(t),y_{k}^{\mathrm{T}}(t),0)$ and $\mathbf{q}_{k}^{\mathrm{R}}(t)=(x_{k}^{\mathrm{R}}(t),y_{k}^{\mathrm{R}}(t),0)$, respectively. The system operates over a time slot $T$, during which user mobility leads to time-varying channel conditions.

\begin{table}[!t]
\centering
\caption{Main Notation}
\label{tab:nomenclature}
\footnotesize
\renewcommand{\arraystretch}{1.15}
\setlength{\tabcolsep}{3pt}
\begin{tabular}{p{1.35cm} p{5.7cm}}
\toprule
\textbf{Notation} & \textbf{Definition} \\
\midrule

$K$ & Number of users \\
$M$ & Number of BS antennas \\
$N$ & Number of STAR-RIS elements \\
$t$ & Time-slot index \\
$T$ & Number of time slots \\
$\Delta t$ & Time-slot duration \\
$\mathbf{q}_{\mathrm{B}}$ & BS position vector \\
$\mathbf{q}_{\mathrm{U}}(t)$ & UAV position vector at time slot $t$ \\
$\mathbf{x}(t)$ & BS transmitted signal vector \\
$\mathbf{w}_{k}(t)$ & Beamforming vector for user $k$ \\
$\mathbf{W}(t)$ & Beamforming matrix \\
$\mathbf{Q}(t)$ & Transmit covariance matrix \\
$\mathbf{G}(t)$ & BS-STAR-RIS channel matrix \\
$\mathbf{h}_{x,k}(t)$ & STAR-RIS-user channel, $x\in\{\mathrm{R},\mathrm{T}\}$ \\
$\boldsymbol{\Theta}_{x}(t)$ & STAR-RIS coefficient matrix \\
$\rho(t)$ & RF power-splitting factor \\
$\beta_{x,n}(t)$ & STAR-RIS amplitude coefficient \\
$\phi_{x,n}(t)$ & STAR-RIS phase shift \\
$\mathbf{h}^{x}_{\mathrm{eq},k}(t)$ & Equivalent cascaded channel \\
$\Gamma_k^x(t)$ & Achievable throughput of user $k$ \\
$\Gamma_{\mathrm{sum}}(t)$ & Total system throughput \\
$P_{\mathrm{in}}(t)$ & Incident RF power at the STAR-RIS \\
$P_{\mathrm{EH}}(t)$ & Harvested RF power \\
$E_{\mathrm{EH}}(t)$ & Harvested RF energy \\
$P_{\mathrm{prop}}(t)$ & UAV propulsion power \\
$P_{\mathrm{tx}}(t)$ & BS transmit power \\
$P_{\mathrm{net}}(t)$ & Net system power consumption \\
$\mathrm{EE}(t)$ & Energy efficiency \\
$V_{\max}$ & Maximum UAV speed \\
$\lambda_1,\lambda_2,\lambda_3$ & Objective weighting coefficients \\
$s_t$ & Environment state at time slot $t$ \\
$a_t$ & Hybrid action vector \\
$a_t^d$ & Discrete UAV movement action \\
$a_t^c$ & Continuous communication-control action vector \\
$V_{\mathrm{QoS}}(t)$ & Aggregate QoS violation \\
$w_{\mathrm{Rate}}$ & Sum-rate reward weight \\
$w_{\mathrm{EE}}$ & Energy-efficiency reward weight \\
$w_{\mathrm{EH}}$ & EH reward weight \\
$\pi_d(a_t^d|s_t)$ & Discrete UAV movement policy \\
$\pi_c(a_t^c|s_t)$ & Continuous control policy \\
$V(s_t)$ & State-value function \\
\bottomrule
\end{tabular}
\end{table}

\subsection{BS Signal Model}

Linear precoding is adopted at the BS. At time slot $t$, the transmitted signal vector is expressed as
\begin{equation}
\mathbf{x}(t)=\sum_{k\in\mathcal{K}}\mathbf{w}_{k}(t)s_{k}(t),
\label{eq:xt}
\end{equation}
where $\mathbf{w}_{k}(t)\in\mathbb{C}^{M\times 1}$ denotes the active beamforming vector associated with user $k$, and $s_{k}(t)$ is the information symbol intended for user $k$. The information symbols are assumed to be independent and identically distributed complex Gaussian random variables, $s_{k}(t)\sim \mathcal{CN}(0,1)$.
Accordingly, the covariance of the transmit signal is determined by the beamforming matrix, and the total BS transmit power at slot $t$ is given by
\begin{equation}
\mathbb{E}\left[\mathbf{x}^{H}(t)\mathbf{x}(t)\right]
=\sum_{k\in\mathcal{K}}\|\mathbf{w}_{k}(t)\|^{2}\leq P_{\max},
\label{eq:Pmax}
\end{equation}
where $P_{\max}$ denotes the maximum available BS transmit power budget. For notational convenience, define the beamforming matrix as $\mathbf{W}(t)=\left[\mathbf{w}_{1}(t),\mathbf{w}_{2}(t),\ldots,\mathbf{w}_{K}(t)\right]\in\mathbb{C}^{M\times K}$. Then, the transmit covariance matrix can be written as $\mathbf{Q}(t)=\mathbf{W}(t)\mathbf{W}^{H}(t)$.

\subsection{Channel Model}
\subsubsection{BS to STAR-RIS}

Let $\mathbf{G}(t)\in\mathbb{C}^{N\times M}$ denote the channel matrix from the BS to the STAR-RIS. Also let the Euclidean distance between the BS and the STAR-RIS at time slot $t$ be given by $d_{\mathrm{BU}}(t)=\|\mathbf{q}_{\mathrm{U}}(t)-\mathbf{q}_{\mathrm{B}}\|_{2}$, while the corresponding elevation angle be defined as $\vartheta(t)=\frac{180}{\pi}\sin^{-1}\!\left(\frac{H_{\mathrm{U}}-H_{\mathrm{B}}}{d_{\mathrm{BU}}(t)}\right)$. Following the probabilistic air-to-ground propagation model~\cite{10051712}, the LoS probability is expressed as $P_{\mathrm{LoS}}(t)=\frac{1}{1+A\exp\!\left(A B-B\vartheta(t)\right)}$, and the NLoS probability is $P_{\mathrm{NLoS}}(t)=1-P_{\mathrm{LoS}}(t)$. Accordingly, the large-scale fading coefficient is given by
\begin{equation}
g_{\mathrm{BU}}(t)=
\left(P_{\mathrm{LoS}}(t)+P_{\mathrm{NLoS}}(t)\varphi\right)
d_{\mathrm{BU}}(t)^{-\alpha},
\end{equation}
where $\alpha$ denotes the path-loss exponent and $\varphi$ accounts for the additional attenuation under NLoS propagation. Let $\widetilde{\mathbf{G}}(t)\in\mathbb{C}^{N\times M}$ denote the small-scale fading matrix, whose entries follow independent circularly symmetric complex Gaussian distributions. The overall BS-STAR-RIS channel is therefore modeled as
\begin{equation}
\mathbf{G}(t)=g_{\mathrm{BU}}(t)\widetilde{\mathbf{G}}(t).
\end{equation}

It is worth noting that the UAV position directly affects both $d_{\mathrm{BU}}(t)$ and $\vartheta(t)$, which in turn determines the large-scale fading coefficient $g_{\mathrm{BU}}(t)$. Consequently, the UAV trajectory has a direct impact on the instantaneous quality of the BS-STAR-RIS link.

\subsubsection{STAR-RIS to UT}

The channels from the STAR-RIS to the users are modeled independently for transmission-side and reflection-side users. Specifically, let $\mathbf{h}_{\mathrm{T},k}(t)\in\mathbb{C}^{N\times 1}$ for $k\in\mathcal{K}_{\mathrm{T}}$ denote the channel from the STAR-RIS to transmission-side user $k$, and $\mathbf{h}_{\mathrm{R},k}(t)\in\mathbb{C}^{N\times 1}$ for $k\in\mathcal{K}_{\mathrm{R}}$ denote the channel from the STAR-RIS to reflection-side user $k$. Also the Euclidean distance between the STAR-RIS and user $k$ can be expressed as $d_{k}(t)=\left\|\mathbf{q}_{k}(t)-\mathbf{q}_{\mathrm{U}}(t)\right\|_{2}$. In this case, the large-scale path loss can be written as $PL_{k}(t)=\kappa d_{k}(t)^{-\hat{\alpha}}$, where $\kappa$ is the path-loss coefficient at the reference distance and $\hat{\alpha}$ is the path-loss exponent. The small-scale fading is modeled using a Rician distribution, which captures both the dominant deterministic propagation component and the scattered random component. Accordingly, the channel vector is expressed as
\begin{align}
\mathbf{h}_{k}(t)
&=
\sqrt{PL_{k}(t)}
\left(
\sqrt{\frac{K_{\mathrm{Ri}}}{1+K_{\mathrm{Ri}}}}
\,\mathbf{h}_{k}^{\mathrm{LoS}}(t)
\right.
\nonumber\\
&\quad\left.
+
\sqrt{\frac{1}{1+K_{\mathrm{Ri}}}}
\,\mathbf{h}_{k}^{\mathrm{NLoS}}(t)
\right),
\label{eq:hk}
\end{align}
where $K_{\mathrm{Ri}}$ denotes the Rician factor. The LoS component is modeled as $\mathbf{h}_{k}^{\mathrm{LoS}}(t)=\mathbf{1}_{N}$, where $\mathbf{1}_{N}\in\mathbb{R}^{N\times 1}$ is the all-ones vector, and the NLoS component follows $\mathbf{h}_{k}^{\mathrm{NLoS}}(t)\sim \mathcal{CN}(\mathbf{0},\mathbf{I}_{N})$.

This model captures both the deterministic LoS contribution and the random scattering effects, with their relative strengths governed by the Rician factor. A larger value of $K_{\mathrm{Ri}}$ indicates a stronger LoS component, whereas a smaller value implies a channel dominated by scattering.


\subsection{Transmission and EH Model}
The STAR-RIS simultaneously supports signal reflection and transmission while allocating a portion of the incident RF energy for harvesting. Let $\rho(t)\in[0,\rho_{\max}]$ denote the power-splitting coefficient at time slot $t$, where $\rho(t)$ is used for EH and $1-\rho(t)$ for communications.

For $x \in \{\mathrm{R}, \mathrm{T}\}$ representing reflection and transmission, respectively, the STAR-RIS coefficient matrix is defined as
\begin{equation}
\boldsymbol{\Theta}_{x}(t)=
\mathrm{diag}\left(\theta_{x,1}(t),\ldots,\theta_{x,N}(t)\right),
\end{equation}
where $\theta_{x,n}(t)=
\sqrt{1-\rho(t)}\,\beta_{x,n}(t)e^{j\phi_{x,n}(t)}.$ The amplitude coefficients satisfy the energy conservation constraint $\beta_{\mathrm{R},n}^{2}(t)+\beta_{\mathrm{T},n}^{2}(t)=1,\quad \forall n,t.$ This model shows that increasing $\rho(t)$ improves EH while reducing the effective communication gain, revealing a fundamental trade-off.
In addition to facilitating signal transmission and reflection, the STAR-RIS also harvests RF energy from the incident BS signal. Based on the transmit covariance matrix $\mathbf{Q}(t)$, the total incident RF power impinging on the STAR-RIS is expressed as
\begin{equation}
P_{\mathrm{in}}(t)=
\sum_{n=1}^{N}\mathbf{g}_{n}^{H}(t)\mathbf{Q}(t)\mathbf{g}_{n}(t),
\label{eq:Pin}
\end{equation}
where $\mathbf{g}_{n}(t)$ denotes the channel vector from the BS antenna array to the $n$-th STAR-RIS element.

The harvested RF power is then given by
\begin{equation}
P_{\mathrm{EH}}(t)=\eta_{\mathrm{RF}}\rho(t)P_{\mathrm{in}}(t),
\label{eq:PEH_RF}
\end{equation}
where $\eta_{\mathrm{RF}}\in(0,1]$ is the RF to direct current energy conversion efficiency. The harvested RF energy during one time slot of duration $\Delta t$ is therefore
\begin{equation}
E_{\mathrm{EH}}(t)=P_{\mathrm{EH}}(t)\Delta t.
\label{eq:EEH_RF}
\end{equation}

From \eqref{eq:PEH_RF}, we can see that the harvested energy increases with both the incident RF power and the power-splitting ratio $\rho(t)$. However, since the same coefficient $\rho(t)$ reduces the effective communication amplitude of the STAR-RIS, maximizing EH alone may degrade the achievable throughput. This highlights the intrinsic trade-off between communication performance and EH.

\subsection{UAV Mobility and Propulsion Energy Model}

The UAV moves in the horizontal plane while maintaining a fixed altitude $H_{\mathrm{U}}$. Let $\Delta t$ denote the slot duration. The UAV motion is constrained by the maximum speed $V_{\max}$ as
\begin{equation}
\|\mathbf{q}_{\mathrm{U}}(t+1)-\mathbf{q}_{\mathrm{U}}(t)\|_{2}\leq V_{\max}\Delta t,\quad \forall t.
\end{equation}

The propulsion power of the rotary-wing UAV is modeled as
\begin{equation}
P_{\mathrm{prop}}(t)=P_0\left(1+\frac{3v^2(t)}{U_{\mathrm{tip}}^2}\right)
+P_i\left(\sqrt{1+\frac{v^4(t)}{4v_0^4}}-\frac{v^2(t)}{2v_0^2}\right)^{\frac{1}{2}},
\label{eq:Pprop}
\end{equation}
where $P_0$ denotes the blade profile power, $P_i$ is the induced power in hovering, $v(t)$ is the UAV speed at time slot $t$, $U_{\mathrm{tip}}$ represents the rotor blade tip speed, and $v_0$ is the mean induced velocity in hovering \cite{8663615}.

The propulsion energy consumed during one time slot is
\begin{equation}
E_{\mathrm{prop}}(t)=P_{\mathrm{prop}}(t)\Delta t.
\label{eq:Eprop}
\end{equation}


For $x \in \{\mathrm{R}, \mathrm{T}\}$ and user $k \in \mathcal{K}_x$, the equivalent cascaded channel is defined as
\begin{equation}
\mathbf{h}_{\mathrm{eq},k}^{x}(t)
=
\mathbf{h}_{x,k}^{H}(t)\boldsymbol{\Theta}_{x}(t)\mathbf{G}(t).
\end{equation}

The received signal is given by
\begin{equation}
\label{eq:received_signal}
y_{k}^{x}(t)
=
\mathbf{h}_{\mathrm{eq},k}^{x}(t)\mathbf{x}(t)+\nu_{k}^{x}(t),
\end{equation}
where $\nu_{k}^{x}(t)\sim\mathcal{CN}(0,\sigma^{2})$.

\section{Performance Metrics and Problem Formulation}
\label{sec:problem_formulation}
\subsection{Performance Metrics}
According to the received signal in \eqref{eq:received_signal}, the signal-to-noise ratio (SNR) is expressed as
\begin{equation}
\mathrm{SNR}_{k}^{x}(t)=
\frac{\left|\mathbf{h}_{\mathrm{eq},k}^{x}(t)\mathbf{w}_{k}(t)\right|^{2}}{\sigma^{2}},
\end{equation}
and the corresponding achievable throughput is
\begin{equation}
\Gamma_{k}^{x}(t)=
B\log_{2}\!\left(1+\mathrm{SNR}_{k}^{x}(t)\right)
\label{eq:GammaSum}
\end{equation}
where $B$ denotes the system bandwidth.

The total system throughput is then given by $\Gamma_{\mathrm{sum}}(t)=
\sum_{x \in \{\mathrm{R},\mathrm{T}\}}
\sum_{k\in\mathcal{K}_{x}}
\Gamma_{k}^{x}(t).
$ The achievable throughput depends jointly on the BS beamforming, STAR-RIS configuration, channel conditions, and UAV position, requiring coordinated optimization. The instantaneous BS transmit power is defined as $
P_{\mathrm{tx}}(t)=\sum_{k\in\mathcal{K}}\|\mathbf{w}_{k}(t)\|^{2}.
\label{eq:Ptx}$
Since the harvested RF power can partially offset the energy expenditure of the system, the net power consumption at slot $t$ is modeled as
\begin{equation}
P_{\mathrm{net}}(t)=P_{\mathrm{tx}}(t)+P_{\mathrm{prop}}(t)-P_{\mathrm{EH}}(t).
\label{eq:Pnet}
\end{equation}
Accordingly, the instantaneous energy efficiency of the considered system is defined as
\begin{equation}
\mathrm{EE}(t)=\frac{\Gamma_{\mathrm{sum}}(t)}{P_{\mathrm{net}}(t)}
=
\frac{\Gamma_{\mathrm{sum}}(t)}{P_{\mathrm{tx}}(t)+P_{\mathrm{prop}}(t)-P_{\mathrm{EH}}(t)}.
\label{eq:EE}
\end{equation}

\subsection{Problem Formulation}
Our optimization problem is to jointly determine the BS beamforming vectors, the STAR-RIS transmission and reflection coefficients, the RF power-splitting factor, and the UAV trajectory that maximize the weighted sum of system throughput, energy efficiency, and harvested energy, while ensuring the fulfillment of the required QoS of all users. To this end, the optimization problem is formulated as
\begin{align}
(\mathrm{P}1):\quad
& \max_{\rho(t),\,\mathbf{W}(t),\,\boldsymbol{\Theta}_{\mathrm{R}}(t),\,\boldsymbol{\Theta}_{\mathrm{T}}(t),\,\mathbf{q}_{\mathrm{U}}(t)}
\nonumber\\
& \sum_{t=1}^{T}
\left[
\lambda_{1}\Gamma_{\mathrm{sum}}(t)
+\lambda_{2}\mathrm{EE}(t)
+\lambda_{3}E_{\mathrm{EH}}(t)
\right]
\label{eq:P1}
\end{align}
subject to
\begin{align}
& \mathrm{C}1:\; \label{eq:C1}
\Gamma_{k}^{\mathrm{R}}(t)\geq \Gamma_{\min},
\quad \forall k\in\mathcal{K}_{\mathrm{R}},\ \forall t,
\\
& \mathrm{C}2:\; \label{eq:C2}
\Gamma_{k}^{\mathrm{T}}(t)\geq \Gamma_{\min},
\quad \forall k\in\mathcal{K}_{\mathrm{T}},\ \forall t,
\\
& \mathrm{C}3:\; \label{eq:C3}
0\leq \rho(t)\leq \rho_{\max},
\quad \forall t,
\\
& \mathrm{C}4:\; \label{eq:C4}
\sum_{k\in\mathcal{K}}\|\mathbf{w}_{k}(t)\|^{2}\leq P_{\max},
\quad \forall t,
\\
& \mathrm{C}5:\; \label{eq:C5}
\beta_{\mathrm{R},n}^{2}(t)+\beta_{\mathrm{T},n}^{2}(t)=1,
\quad \forall n,\ \forall t,
\\
& \mathrm{C}6:\; \label{eq:C6}
0\leq \beta_{\mathrm{R},n}(t)\leq 1, \quad
0\leq \beta_{\mathrm{T},n}(t)\leq 1
\quad \forall n,\ \forall t,
\\
& \mathrm{C}7:\; \label{eq:C7}
0 \leq \phi_{\mathrm{R},n}(t) < 2\pi,\quad
0 \leq \phi_{\mathrm{T},n}(t) < 2\pi,
\: \forall n,\ \forall t,
\\
& \mathrm{C}8:\; \label{eq:C8}
\left\|\mathbf{q}_{\mathrm{U}}(t+1)-\mathbf{q}_{\mathrm{U}}(t)\right\|_{2}
\leq V_{\max}\Delta t,
\quad t=1,\ldots,T-1,
\\
& \mathrm{C}9:\; \label{eq:C9}
x_{\min}\leq x_{\mathrm{U}}(t)\leq x_{\max},
\quad \forall t,
\\
& \mathrm{C}10:\; \label{eq:C10}
y_{\min}\leq y_{\mathrm{U}}(t)\leq y_{\max},
\quad \forall t.
\end{align}

In problem $(\mathrm{P}1)$, $\lambda_{1}$, $\lambda_{2}$, and $\lambda_{3}$ are nonnegative weighting coefficients that balance the trade-off among throughput maximization, energy-efficiency enhancement, and harvested-energy improvement. Constraints \eqref{eq:C1} and \eqref{eq:C2} ensure the minimum throughput requirements of the reflection-side and transmission-side users, respectively. Constraint \eqref{eq:C3} specifies the feasible range of the power-splitting factor, while \eqref{eq:C4} enforces the BS transmit power constraint. Constraints \eqref{eq:C5} and \eqref{eq:C6} define the STAR-RIS amplitude constraints. Constraint \eqref{eq:C7} imposes the feasible range of the STAR-RIS phase shifts. Constraint \eqref{eq:C8} characterizes the UAV mobility constraint across consecutive time slots, and constraints \eqref{eq:C9} and \eqref{eq:C10} enforce the UAV service-region boundaries.

Relation between $(\lambda_1,\lambda_2,\lambda_3)$ and $(w_{\mathrm{Rate}}, w_{\mathrm{EE}}, w_{\mathrm{EH}})$: These two sets of weights serve different purposes. $\lambda_1$, $\lambda_2$, and $\lambda_3$ define the original optimization problem $(\mathrm{P}1)$ they say how much we ultimately care about sum rate, energy efficiency, and harvested energy in the true objective we are trying to solve. $w_{\mathrm{Rate}}$, $w_{\mathrm{EE}}$, and $w_{\mathrm{EH}}$, on the other hand, are used only inside the RL reward function, and they are applied after each metric has been rescaled to the $[0,1]$ range. This rescaling step is necessary: sum rate, EE, and harvested energy naturally live on very different numerical scales, so combining them directly with the $\lambda$ weights would let one objective dominate the training signal.

\section{Proposed Hybrid PPO-Based Methodology}
\label{sec:methodology}

The considered optimization problem is highly nonconvex due to the strong coupling among communication, EH, and mobility variables. In particular, the achievable throughput in \eqref{eq:GammaSum}, the harvested energy in \eqref{eq:EEH_RF}, and the energy efficiency in \eqref{eq:EE} all depend jointly on the beamforming matrix $\mathbf{W}(t)$, the STAR-RIS coefficient matrices $\mathbf{\Theta}_{\mathrm{R}}(t)$ and $\mathbf{\Theta}_{\mathrm{T}}(t)$, the power-splitting ratio $\rho(t)$, and the UAV position $\mathbf{q}_{\mathrm{U}}(t)$. Moreover, the time-varying UT positions and UAV mobility induce time-varying channel conditions, which make conventional iterative optimization methods computationally expensive for online control.
\begin{figure*}[!t]
\centering
\includegraphics[width=0.9\linewidth]{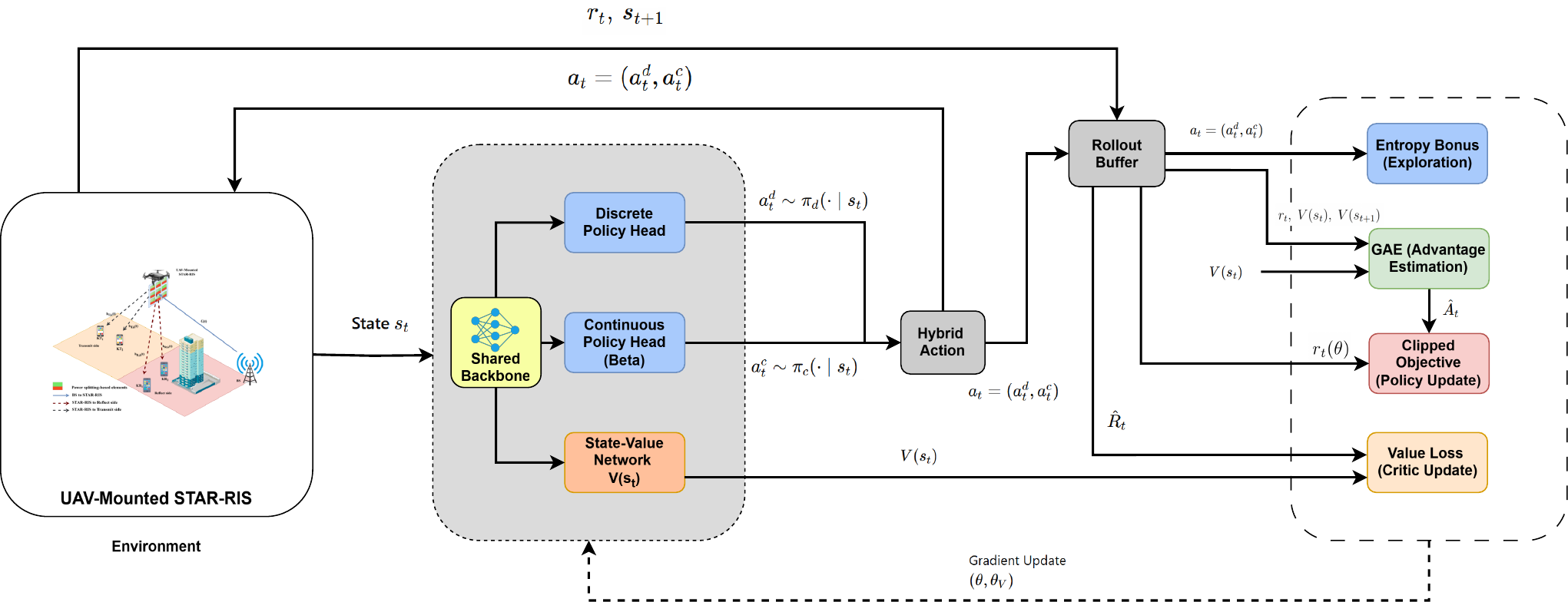}
\caption{Hybrid PPO-based UAV-mounted STAR-RIS control framework.}
\label{fig:hppo_framework}
\end{figure*}

To address this challenge, a hybrid DRL framework based on PPO is developed. The proposed method learns a control policy directly through interaction with the environment, without requiring explicit decomposition of problem $(\mathrm{P}1)$. The control variables are structured into discrete and continuous components: the UAV movement is modeled as a discrete action, whereas the STAR-RIS amplitudes and phases, beamforming parameters, and RF power-splitting factor are treated as continuous variables. Accordingly, a hybrid actor-critic architecture is adopted, where a categorical distribution governs UAV motion and Beta distributions model the continuous control variables. The overall framework follows the clipped-surrogate PPO paradigm in \cite{schulman2017ppo}, and its architecture is illustrated in Fig.~\ref{fig:hppo_framework}.

\subsection{Markov decision process(MDP) Formulation}

We formulate our problem as MDP, defined by the tuple $\mathcal{M}=\left(\mathcal{S}, \mathcal{A}, \mathcal{P}, \mathcal{R}, \gamma\right)$, where $\mathcal{S}$ represents the state space, $\mathcal{A}$ the action space, $\mathcal{P}$ the state transition probability distribution, $\mathcal{R}$ the reward function, and $\gamma\in (0,1)$ the discount factor. At each time slot $t$, the agent observes the current state $s_t\in\mathcal{S}$, executes an action $a_t\in\mathcal{A}$ according to policy $\pi(a_t|s_t)$, receives an immediate reward $r_t$ and the environment transitions to the next state $s_{t+1}$.

\subsubsection{State Space}

The state is designed to capture the instantaneous channel conditions and UAV location. Consistent with the environment implementation, it includes the real and imaginary parts of the BS-STAR-RIS channel $\mathbf{G}(t)$, the BS-user channel terms, the STAR-RIS-user channel terms, and the normalized UAV position. Accordingly, the state at time slot $t$ is represented as
\begin{equation}
\begin{aligned}
s_t = \big[
&\Re\{\mathbf{G}(t)\},
\Im\{\mathbf{G}(t)\},
\Re\{\mathbf{H}_{BK}(t)\},
\Im\{\mathbf{H}_{BK}(t)\}, \\
&\Re\{\mathbf{H}_{RK}(t)\},
\Im\{\mathbf{H}_{RK}(t)\},
x_{\mathrm{R}}(t),
y_{\mathrm{R}}(t),
z_{\mathrm{R}}(t)
\big]
\end{aligned}
\label{eq:state_method}
\end{equation}
where $\mathbf{H}_{BK}(t)$ and $\mathbf{H}_{RK}(t)$ denote the stacked BS-user and STAR-RIS-user channel representations used in the observation vector. This state construction enables the agent to adapt its decisions according to both the propagation environment and the UAV geometry.

\subsubsection{Action Space}

The action is hybrid and consists of a discrete part and a continuous part, i.e., $a_t=\left(a_t^{d},a_t^{c}\right)$. The discrete action $a_t^{d}$ controls the UAV horizontal movement and is selected from $a_t^{d}\in\left\{\text{left},\text{right},\text{forward},\text{backward},\text{stay}\right\}$.
The continuous action vector $a_t^{c}$ represents all communication-control variables, namely
\begin{equation}
a_t^{c}=
\left[
\boldsymbol{\phi}_{\mathrm{R}}(t),\,
\boldsymbol{\beta}_{\mathrm{R}}(t),\,
\boldsymbol{\delta}(t),\,
\boldsymbol{\phi}_{\mathbf{W}}(t),\,
\boldsymbol{\beta}_{\mathbf{W}}(t),\,
\rho(t)
\right],
\label{eq:continuous_action_method}
\end{equation}
where $\boldsymbol{\phi}_{\mathrm{R}}(t)\in\mathbb{R}^{N}$ contains the reflection phases, $\boldsymbol{\beta}_{\mathrm{R}}(t)\in\mathbb{R}^{N}$ is the reflection amplitudes, $\boldsymbol{\delta}(t)\in\mathbb{R}^{N}$ is the element-wise transmission-side phase-switching variables, $\boldsymbol{\phi}_{\mathbf{W}}(t)\in\mathbb{R}^{MK}$ denotes the beamforming phases, $\boldsymbol{\beta}_{\mathbf{W}}(t)\in\mathbb{R}^{MK}$ represents the beamforming amplitudes, and $\rho(t)\in[0,1]$ corresponds to the RF power-splitting factor.

\subsubsection{Reward Function}

For the reward function, the reward is designed to first enforce user QoS and then optimize the communication-energy objectives within the feasible region. Let the aggregate QoS violation at time slot $t$ be $V_{\mathrm{QoS}}(t)=\sum_{k\in\mathcal{K}} \max\left(0,\Gamma_{\min}-\Gamma_k(t)\right)$, where $\Gamma_k(t)$ is the achieved throughput of user $k$ and $\Gamma_{\min}$ is the minimum required throughput. To balance the heterogeneous objectives, namely the sum-rate, energy efficiency, and harvested energy, a min-max normalization strategy is adopted. Such normalization is widely used in DRL-based wireless optimization to improve numerical stability and learning efficiency~\cite{WU2025103744}. The normalized value of a generic metric $X(t)$ is defined as $\bar{X}(t)=\frac{X(t)-X_{\min}}{X_{\max}-X_{\min}}$, where $X_{\min}$ and $X_{\max}$ denote the minimum and maximum values observed during a warm-up phase.
Based on this normalization, the reward is defined as
\begin{equation}
r_t=
\begin{cases}
-\lambda_{\mathrm{Q}}
\tanh\!\left(
\dfrac{V_{\mathrm{QoS}}(t)}{\xi_{\mathrm{Q}}}
\right),
& V_{\mathrm{QoS}}(t)>0,
\\[1ex]
\begin{split}
w_{\mathrm{Rate}}\,
\bar{\Gamma}_{\mathrm{sum}}(t)
+ w_{\mathrm{EE}}\,
\bar{\mathrm{EE}}(t)
\\
+ w_{\mathrm{EH}}\,
\bar{E}_{\mathrm{EH}}(t),
\end{split}
& V_{\mathrm{QoS}}(t)=0,
\end{cases}
\label{eq:reward_method}
\end{equation}
where $\bar{\Gamma}_{\mathrm{sum}}(t)$, $\bar{\mathrm{EE}}(t)$, and $\bar{E}_{\mathrm{EH}}(t)$ are the normalized sum-rate, energy efficiency, and harvested energy, respectively. The parameters $\lambda_{\mathrm{Q}}$ and $\xi_{\mathrm{Q}}$ control the magnitude and smoothness of the QoS penalty with $\xi_{\mathrm{Q}}$ set to $K \, \Gamma_{\min}$, while $w_{Rate}$, $w_{ee}$, and $w_{EH}$ determine the relative importance of the three objectives. This reward formulation reflects the structure of problem $(\mathrm{P}1)$: infeasible actions are penalized through a bounded and smooth QoS violation term, whereas feasible actions are evaluated based on a normalized multi-objective performance metric. This design improves training stability and facilitates balanced learning across objectives with different scales.

\subsection{Hybrid Actor-Critic Network Architecture}

To solve the formulated MDP, a hybrid actor-critic architecture is employed, which is designed to handle both discrete and continuous control variables in a unified framework. The network consists of a shared feature extractor followed by separate branches for policy learning and value estimation.

Specifically, given the input state $s_t$, a multilayer perceptron (MLP) with three fully connected layers and $\tanh$ activation is first used to extract a latent representation, denoted by $\mathbf{z}_t = f_{\mathrm{enc}}(s_t;\theta_{\mathrm{enc}})$, where $\theta_{\mathrm{enc}}$ represents the encoder parameters. This shared representation captures the essential features of the environment and serves as a common input to both the policy and value networks.

Based on this latent representation, the policy is decomposed into two components to accommodate the hybrid action space. The discrete component models the UAV movement using a categorical distribution. Specifically, the logits are generated as $\boldsymbol{\ell}_t = f_d(\mathbf{z}_t;\theta_d)$, yielding the policy $\pi_d(a_t^d|s_t) = \mathrm{Categorical}(\boldsymbol{\ell}_t)$. 

In parallel, the continuous component models the communication-related control variables using independent Beta distributions. The corresponding shape parameters are obtained as $\boldsymbol{\alpha}_t^{\mathrm{B}} = \mathrm{Softplus}(f_{\alpha}(\mathbf{z}_t;\theta_{\alpha})) + 1$ and $\boldsymbol{\zeta}_t^{\mathrm{B}} = \mathrm{Softplus}(f_{\zeta}(\mathbf{z}_t;\theta_{\zeta})) + 1$. The resulting policy is expressed as $\pi_c(a_t^c|s_t) = \prod_{i=1}^{D_c} \mathrm{Beta}(a_{t,i}^c; \alpha_{t,i}^{\mathrm{B}}, \zeta_{t,i}^{\mathrm{B}})$, where $D_c$ denotes the dimension of the continuous action vector. This formulation ensures that all continuous actions remain within the bounded interval $[0,1]$.

In addition to the policy network, a value network is used to estimate the state-value function, defined as $V(s_t) = f_V(\mathbf{z}_t;\theta_V)$. This component provides a baseline for variance reduction and guides the policy update process.

Finally, the overall policy is factorized as $\pi(a_t|s_t) = \pi_d(a_t^d|s_t)\,\pi_c(a_t^c|s_t)$. Accordingly, the log-probability decomposes as $\log \pi(a_t|s_t) = \log \pi_d(a_t^d|s_t) + \sum_{i=1}^{D_c} \log \pi_{c,i}(a_{t,i}^c|s_t)$, and the total entropy is computed as the sum of the discrete and continuous entropy terms. This hybrid formulation enables efficient and coordinated optimization of both UAV mobility and communication parameters.

\subsection{Rollout Collection}

During training, the agent interacts with the environment for a fixed rollout horizon $T_r$. At each time slot $t$, the discrete UAV action is sampled as $a_t^d\sim\pi_d(\cdot|s_t)$ and the continuous action vector as $a_t^c\sim\pi_c(\cdot|s_t)$. The hybrid action $a_t=(a_t^d,a_t^c)$ is then executed in the environment, and the resulting transition
\begin{equation}
\left(s_t,a_t^d,a_t^c,r_t,s_{t+1},\log\pi(a_t|s_t),V(s_t)\right),
\label{eq:transition_method}
\end{equation}
is stored in the rollout buffer. The rollout-based data collection follows the standard PPO paradigm of alternating between environment interaction and policy optimization over the collected on-policy samples \cite{schulman2017ppo}.

\subsection{Generalized Advantage Estimation}

After collecting one rollout, generalized advantage estimation (GAE) is used to compute the advantage values. The temporal-difference residual at time slot $t$ is
\begin{equation}
\delta_t=r_t+\gamma V(s_{t+1})-V(s_t),
\label{eq:delta_method}
\end{equation}
where the advantage estimate is recursively computed as $
\hat{A}_t=\delta_t+\gamma\lambda\hat{A}_{t+1},$ and $\lambda$ is the GAE parameter. The return target is then obtained as $\hat{R}_t=\hat{A}_t+V(s_t).$ To improve optimization stability, the advantages are normalized within each rollout according to
\begin{equation}
\hat{A}_t\leftarrow\frac{\hat{A}_t-\mu_A}{\sigma_A},
\label{eq:adv_norm_method}
\end{equation}
where $\mu_A$ and $\sigma_A$ are the sample mean and standard deviation of the advantages.

\subsection{PPO-Based Policy Update}

Let $\pi_{\theta_{\mathrm{old}}}$ denote the behavior policy that generated the rollout and $\pi_{\theta}$ the updated policy. The probability ratio is defined as
\begin{equation}
r_t(\theta)=\frac{\pi_{\theta}(a_t|s_t)}{\pi_{\theta_{\mathrm{old}}}(a_t|s_t)}
=\exp\!\Big(\log\pi_{\theta}(a_t|s_t)-\log\pi_{\theta_{\mathrm{old}}}(a_t|s_t)\Big).
\label{eq:ratio_method}
\end{equation}

The PPO clipped surrogate objective is
\begin{equation}
\mathcal{L}_{\mathrm{clip}}(\theta)=
\mathbb{E}_t\!\left[
\min\Big(
r_t(\theta)\hat{A}_t,\,
\mathrm{clip}(r_t(\theta),1-\epsilon,1+\epsilon)\hat{A}_t
\Big)
\right],
\label{eq:ppo_clip_method}
\end{equation}
where $\epsilon$ is the clipping coefficient. This objective prevents excessively large policy updates while preserving sufficient learning progress \cite{schulman2017ppo}.

The value-function loss is
\begin{equation}
\mathcal{L}_{V}(\theta_V)=
\frac{1}{2}\mathbb{E}_t\!\left[\left(\hat{R}_t-V_{\theta_V}(s_t)\right)^2\right],
\label{eq:value_loss_method}
\end{equation}
and the entropy regularization term is
\begin{equation}
\mathcal{L}_{\mathrm{ent}}(\theta)=
\mathbb{E}_t\!\left[\mathcal{H}\big(\pi_{\theta}(\cdot|s_t)\big)\right].
\label{eq:entropy_method}
\end{equation}

The final loss minimized during training is
\begin{equation}
\mathcal{L}(\theta,\theta_V)=
-\mathcal{L}_{\mathrm{clip}}(\theta)
+c_1\mathcal{L}_{V}(\theta_V)
-c_2\mathcal{L}_{\mathrm{ent}}(\theta),
\label{eq:total_loss_method}
\end{equation}
where $c_1$ and $c_2$ are the coefficients of the value loss and entropy bonus, respectively.



\begin{algorithm}[t]
\caption{Hybrid PPO-Based Training for UAV-Mounted STAR-RIS Control}
\label{alg:hppo}
\begin{algorithmic}[1]
\State Initialize environment, hybrid actor-critic network, and rollout buffer
\For{each training iteration}
    \State Observe current state $s_t$
    \For{$t=1$ to $T_r$}
        \State Sample discrete action $a_t^d \sim \pi_d(\cdot|s_t)$
        \State Sample continuous action $a_t^c \sim \pi_c(\cdot|s_t)$
        \State Execute $a_t=(a_t^d,a_t^c)$ in the environment
        \State Observe reward $r_t$, next state $s_{t+1}$, and done flag $d_t$
        \State Store transition in the rollout buffer
        \State Set $s_t\gets s_{t+1}$
    \EndFor
    \State Compute $\hat{A}_t$ and $\hat{R}_t$ using GAE
    \State Normalize the advantages
    \For{each update epoch}
        \For{each mini-batch}
            \State Recompute log-probabilities, entropy, and values
            \State Compute the clipped PPO objective
            \State Compute the value and entropy terms
            \State Update the network parameters using Adam
            \State Clip gradients
        \EndFor
        \If{approximate KL divergence exceeds the threshold}
            \State Break
        \EndIf
    \EndFor
\EndFor
\State Output the trained hybrid policy
\end{algorithmic}
\end{algorithm}

\subsection{Computational Complexity }
Here, we analyze the computational complexity of the proposed H-PPO algorithm using the big-$O$ notation. Let $d_s$, $d_c$, and $d_{dis}$ denote the dimensions of the state space, the continuous action space, and the discrete UAV movement space, respectively. The H-PPO network consists of a shared feature extractor, a categorical policy head for UAV movement, Beta-distribution policy heads for the continuous control variables, and a value head. Therefore, the per-step forward complexity is given by
\begin{equation}
O_{\mathrm{H\text{-}PPO}} =
O\left(
d_s n_1 + n_1 n_2 + n_2 n_3 + n_3(d_{dis} + 2d_c + 1)
\right),
\end{equation}
where $n_1$, $n_2$, and $n_3$ denote the hidden-layer sizes. For each training iteration, the agent collects a rollout of length $T_r$ and updates the policy over $E_{\mathrm{PPO}}$ epochs. Thus, the total training complexity over $N_{\mathrm{iter}}$ iterations is expressed as
\begin{align}
O_{\mathrm{train}}
&=
O\Bigg(
N_{\mathrm{iter}}(1+E_{\mathrm{PPO}})T_r
\nonumber\\
&\quad \times
\Big[
d_s n_1 + n_1 n_2 + n_2 n_3
+ n_3(d_{dis} + 2d_c + 1)
\Big]
\Bigg).
\end{align}
This complexity indicates that the additional cost of H-PPO mainly arises from the hybrid policy structure, namely the categorical head for discrete UAV movement and the Beta-distribution heads for continuous control. Since this cost scales linearly with $d_{dis}$ and $d_c$, the proposed H-PPO framework remains computationally feasible. During deployment, only one forward pass through the trained policy is required, whose complexity is given by
\begin{equation}
O_{\mathrm{deploy}} =
O\left(
d_s n_1 + n_1 n_2 + n_2 n_3
+ n_3(d_{dis} + 2d_c)
\right).
\end{equation}
Therefore, the trained H-PPO policy is suitable for online UAV-mounted STAR-RIS control.

\section{Simulation Results and Discussion}
\label{sec:results}

In this section, we investigate the convergence performance of the H-PPO model against standard DRL methods, namely PPO and TD3. Then, we evaluate the impact of key system parameters on the proposed framework, including the minimum rate requirement, the number of STAR-RIS elements, the BS transmit power, and the number of users. Finally, we examine the effect of the reward-weighting strategy on the trade-off among energy efficiency, sum rate, and harvested energy. The simulation parameters are summarized in Table~\ref{tab:sim_params}, while Table~\ref{tab:hyperparams} illustrates the DRL algorithms' hyperparameters used. All DRL agents are trained for $10^6$ interaction steps. 

\subsection{Baseline Algorithms}
\label{sec:baselines}
The proposed H-PPO framework is benchmarked against two widely used DRL baselines, chosen to isolate the specific contribution of the hybrid action-space design:
\begin{itemize}
\item {PPO} \cite{schulman2017ppo}: a clipped-surrogate on-policy actor-critic method that, in its conventional form, treats the entire action space as continuous. To apply standard PPO to our problem, the discrete UAV movement is relaxed into a continuous variable that is subsequently rounded to the nearest valid direction. PPO is selected as a baseline because it shares the same underlying clipped-objective optimizer and hyperparameter family as H-PPO (see Table~\ref{tab:hyperparams}), so that any performance gap can be attributed specifically to the hybrid discrete/continuous action-space design rather than to differences in the optimization algorithm itself.
\item {TD3} \cite{10704600}: an off-policy, deterministic actor-critic method that uses twin critics and delayed policy updates to mitigate value overestimation. TD3 is included as a baseline because it represents a structurally different (off-policy, deterministic, replay-buffer-based) class of DRL algorithm than PPO/H-PPO, allowing the evaluation to cover both on-policy and off-policy paradigms. Similar to PPO, the discrete UAV action is handled by relaxing it to a continuous surrogate.
\end{itemize}
Both baselines are tuned using the hyperparameters listed in Table~\ref{tab:hyperparams}, which follow standard settings reported in the DRL literature for wireless resource-allocation tasks.

\begin{table}[!t]
\centering
\caption{Simulation Parameters}
\label{tab:sim_params}
\renewcommand{\arraystretch}{1.2}
\setlength{\tabcolsep}{6pt}
\begin{tabular}{l l l}
\toprule
\textbf{Symbol} & \textbf{Description} & \textbf{Value} \\
\midrule
$K_R, K_T$ & Reflect / Transmit users & $1 / 1$ \\
$M$ & BS antennas & $4$ \\
$N$ & STAR-RIS elements & $60$ \\
$T$ & Episode length & $100$ \\
$B$ & Bandwidth & $1$ MHz \\
$P_{\max}$ & Max transmit power & $1$ W \\
$\sigma^2$ & Noise power & $4.00 \times 10^{-15}$ \\
$a, b$ & LoS environment constants & $9.61,\; 0.16$ \\
$\alpha$ & Path-loss exponent (BS–RIS) & $2.2$ \\
$\hat{\alpha}$ & Path-loss exponent (RIS–UT) & $2.5$ \\
$\kappa$ & Reference path-loss coefficient & $10^{-3}$ \\
$K_R^{\text{ric}}$ & Rician factor & $20$ \\
$V_{\max}$ & Max UAV speed & $10$ m/s \\
$z_{\mathrm{uav}}$ & UAV altitude & $30$ m \\
$P_{\mathrm{hover}}$ & Hover power & $80$ W \\
$k_v$ & Propulsion coefficient & $2$ \\
$R_{\min}$ & Rate threshold & $15$ Mbps \\
$\rho$ & Power splitting ratio & $[0, 0.9]$ \\
$w_{Rate}, w_{EE}, w_{EH}$ & Reward weights & $1,\;1,\;1$ \\
\bottomrule
\end{tabular}
\end{table}

\begin{table}[!t]
\centering
\caption{DRL Algorithms Hyperparameters}
\label{tab:hyperparams}
\renewcommand{\arraystretch}{1.2}
\setlength{\tabcolsep}{6pt}
\begin{tabular}{l c c c}
\toprule
\textbf{Parameter} & \textbf{PPO} & \textbf{Hybrid PPO} & \textbf{TD3} \\
\midrule
Learning rate & $3\times10^{-4}$ & $3\times10^{-4}$ & $3\times10^{-5}$ \\
Rollout steps & $2048$ & $2048$ & -- \\
Batch size & $256$ & $256$ & $256$ \\
Epochs & $10$ & $10$ & -- \\
Discount factor $\gamma$ & $0.99$ & $0.99$ & $0.99$ \\
GAE parameter $\lambda$ & $0.95$ & $0.95$ & -- \\
Clip coefficient $\epsilon$ & $0.2$ & $0.2$ & -- \\
Entropy coef. & $0.0003$ & $0.0003$ & -- \\
Value loss coef. & $0.5$ & $0.5$ & -- \\
Max grad norm & $0.5$ & $0.5$ & -- \\
Target KL & $0.03$ & $0.03$ & -- \\
Replay buffer size & -- & -- & $10^6$ \\
Policy delay & -- & -- & $2$ \\
Target noise std & -- & -- & $0.2$ \\
Noise clip & -- & -- & $0.5$ \\
\bottomrule
\end{tabular}
\end{table}

\begin{figure}[H]
\centering
\includegraphics[width=0.9\linewidth]{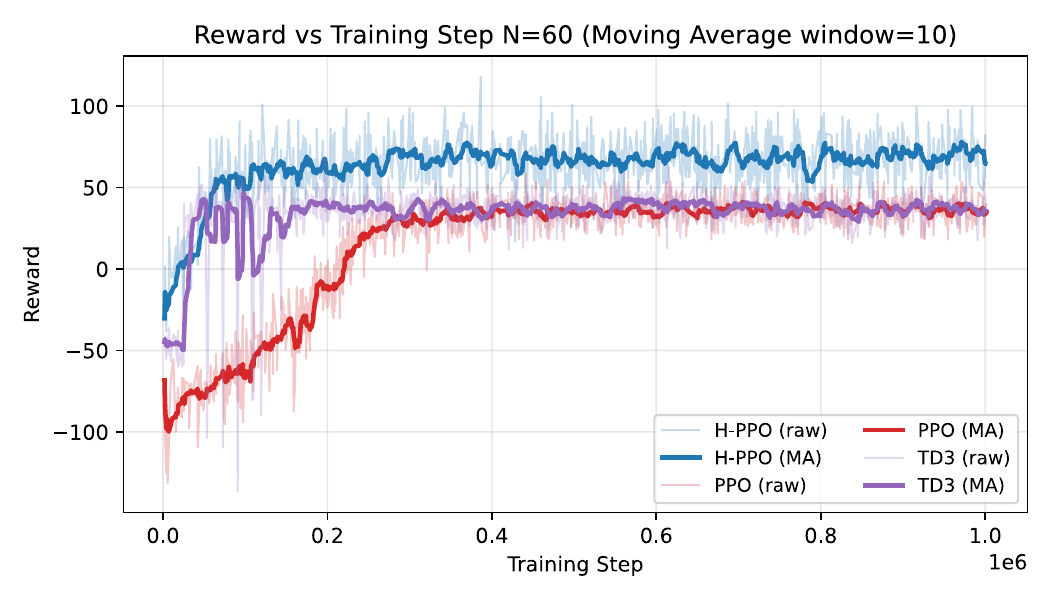}
\caption{Training reward versus training steps for different DRL algorithms.}
\label{fig:reward_training}
\end{figure}

\subsection{Convergence Behavior}

Fig.~\ref{fig:reward_training} compares the training rewards of the proposed H-PPO algorithm with the PPO and TD3 algorithms. The proposed H-PPO achieves significantly faster convergence and improved stability compared to the baseline methods. Specifically, H-PPO exits the infeasible region with negative rewards within approximately $2\times10^{4}$ training steps, reaches a high-reward region around 60 after nearly $6\times10^{4}$ steps, and stabilizes within the range of 65–75 after approximately $10^{5}$ steps. In contrast, PPO remains in the negative-reward region until nearly $2.2\times10^{5}$ steps and only stabilizes at a lower reward level of approximately 30–40 after about $3\times10^{5}$ steps. Although TD3 achieves positive rewards at an early stage, it exhibits pronounced oscillations and does not converge to a stable trajectory until approximately $2.5\times10^{5}$ steps, with a lower final reward level of about 35–45. This behavior is mainly due to the hybrid action space, which enables efficient exploration of both discrete UAV actions and continuous communication-control variables, and to the bounded QoS penalty, which helps avoid unstable reward variations. By contrast, PPO converges more slowly because it does not efficiently explore the hybrid action space, while TD3 suffers from oscillatory behavior due to its deterministic policy structure.


\subsection{Impact of key system parameters}

\begin{figure}[t]
\centering
\subfloat[Energy efficiency\label{fig:ee_vs_rmin}]{
    \includegraphics[width=0.9\linewidth]{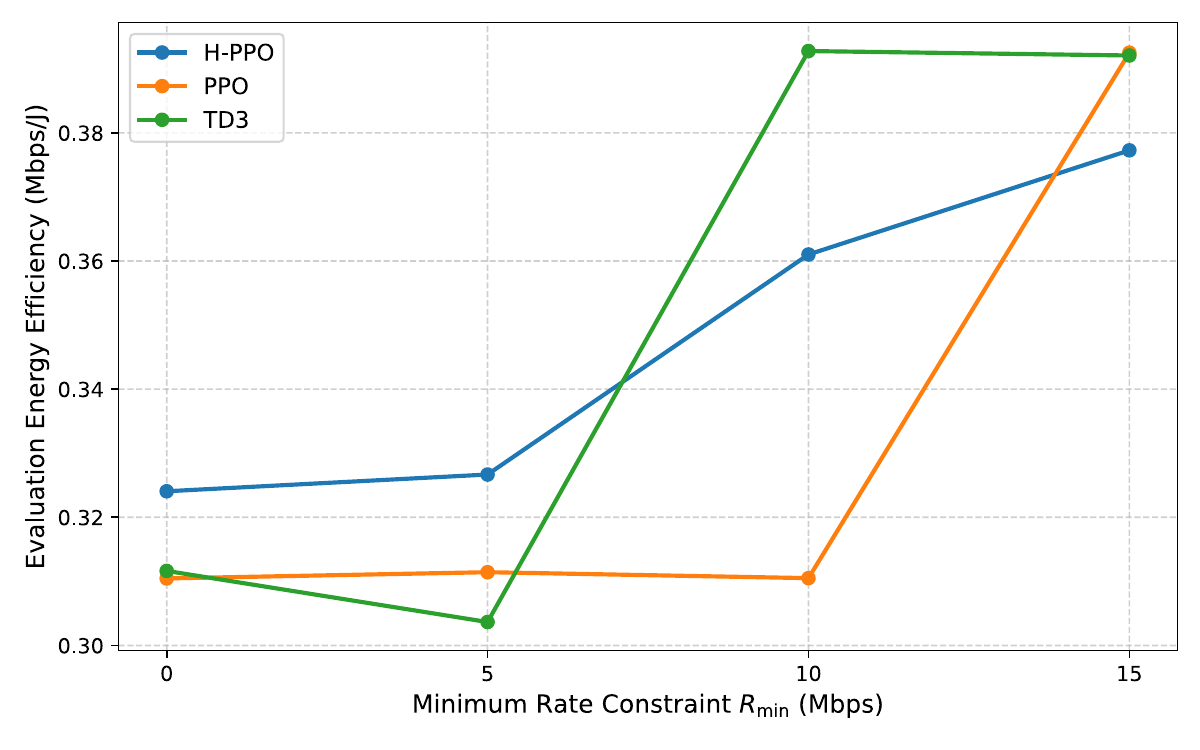}
}
\hfill
\subfloat[Sum-rate\label{fig:sumrate_vs_rmin}]{
    \includegraphics[width=0.9\linewidth]{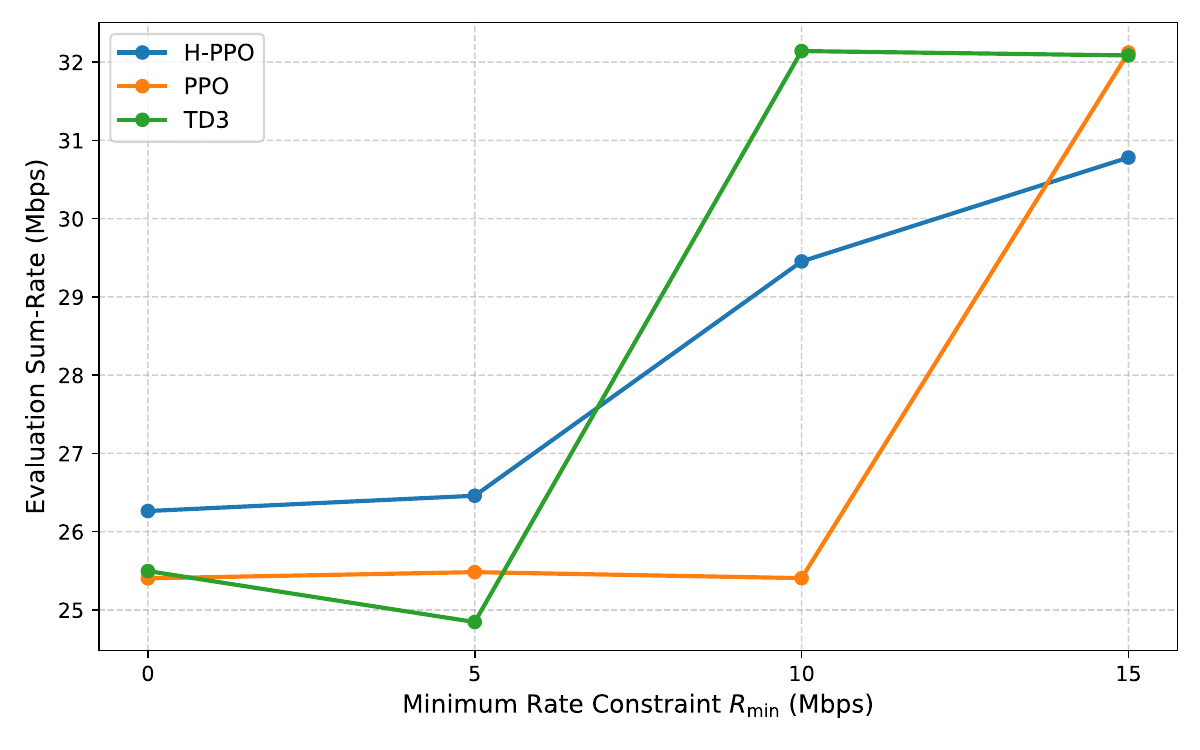}
}
\hfill
\subfloat[Harvested energy\label{fig:eh_vs_rmin}]{
    \includegraphics[width=0.9\linewidth]{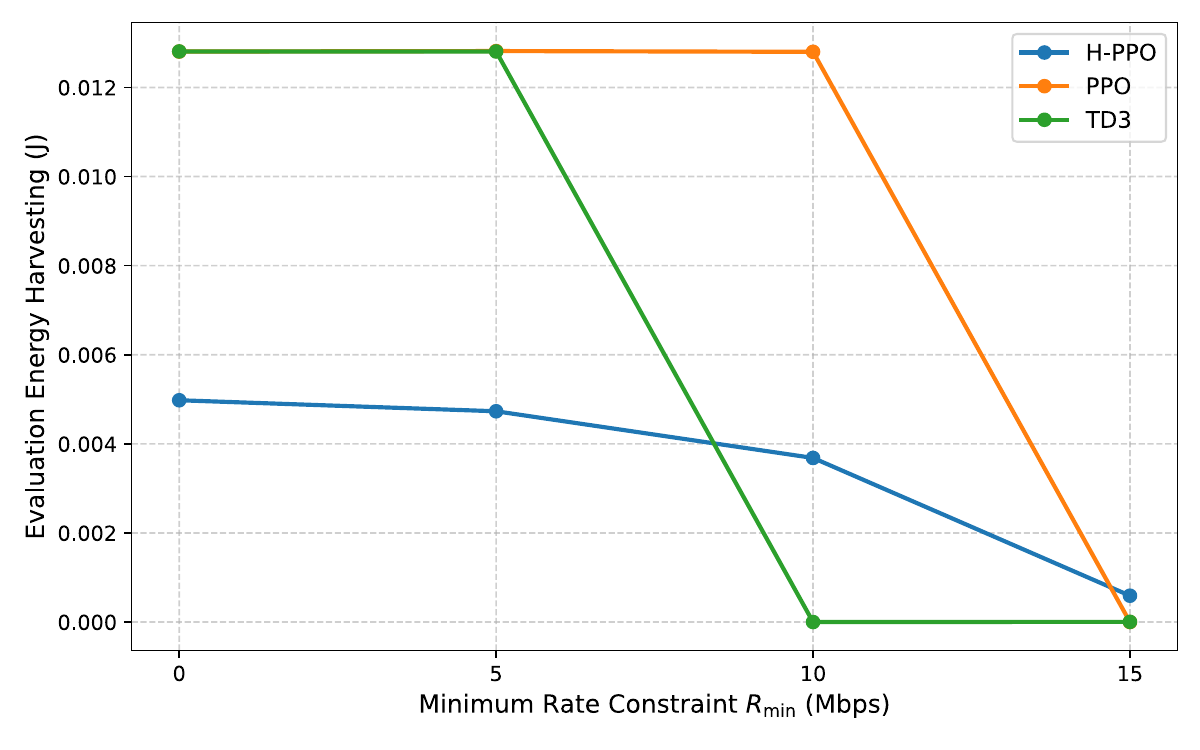}
}
\caption{Performance comparison of the H-PPO, PPO, and TD3 algorithms under different minimum rate constraints $R_{\min}$.}
\label{fig:combined_rmin}
\end{figure}

Fig.~\ref{fig:combined_rmin} illustrates the evaluation performance of the H-PPO, PPO, and TD3 algorithms under different QoS constraints. For all algorithms, as $R_{\min}$ increases, both energy efficiency and sum-rate improve, since the system prioritizes communication performance to satisfy user requirements, which is clear from Fig.~\ref{fig:combined_rmin}(a) and Fig.~\ref{fig:combined_rmin}(b), respectively. However, Fig.~\ref{fig:combined_rmin}(c) shows a decreasing trend in harvested energy, reflecting the trade-off introduced by the power-splitting mechanism. In other words, higher QoS requirements force more power allocation to information transmission, reducing the portion available for EH. Compared to PPO and TD3, H-PPO achieves a smoother performance transition across all QoS levels. TD3 exhibits abrupt performance degradation and EH collapse at high constraints, while PPO shows delayed adaptation. In contrast, H-PPO maintains stable and consistent performance, indicating its ability to effectively balance conflicting objectives.

\begin{figure}[t]
\centering
\subfloat[Energy efficiency\label{fig:ee_vs_N}]{
    \includegraphics[width=0.9\linewidth]{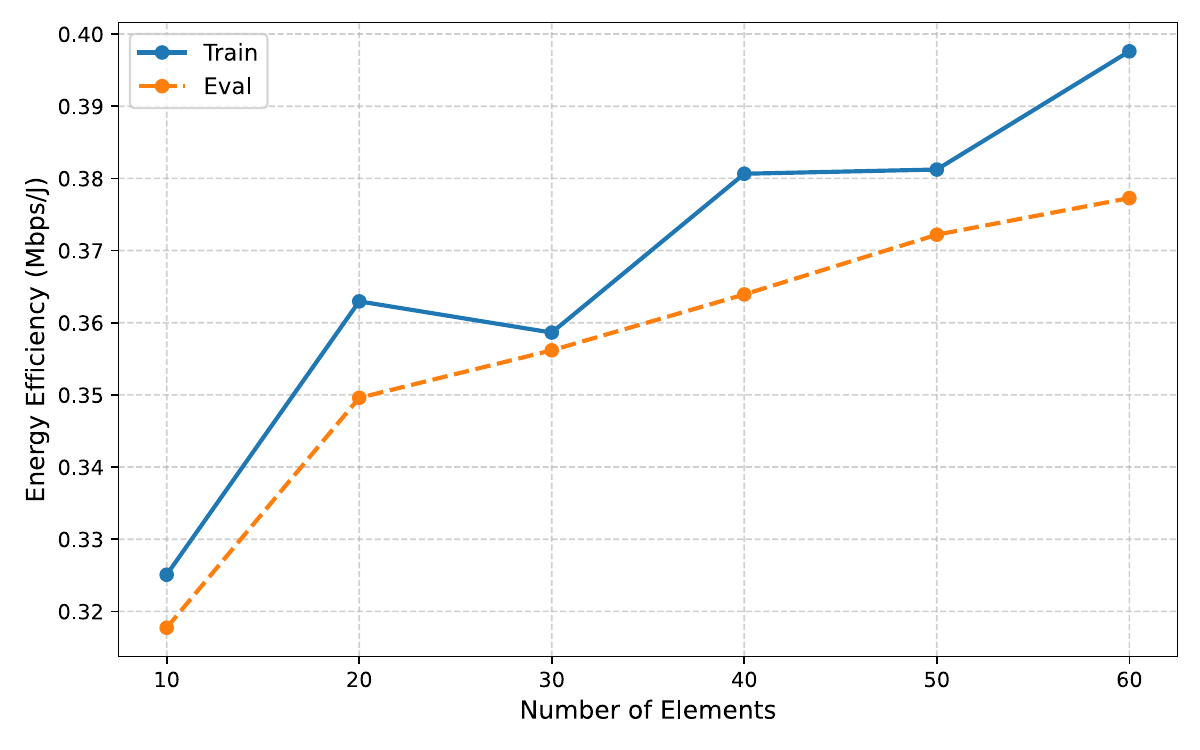}
}
\hfill
\subfloat[Sum-rate\label{fig:sr_vs_N}]{
    \includegraphics[width=0.9\linewidth]{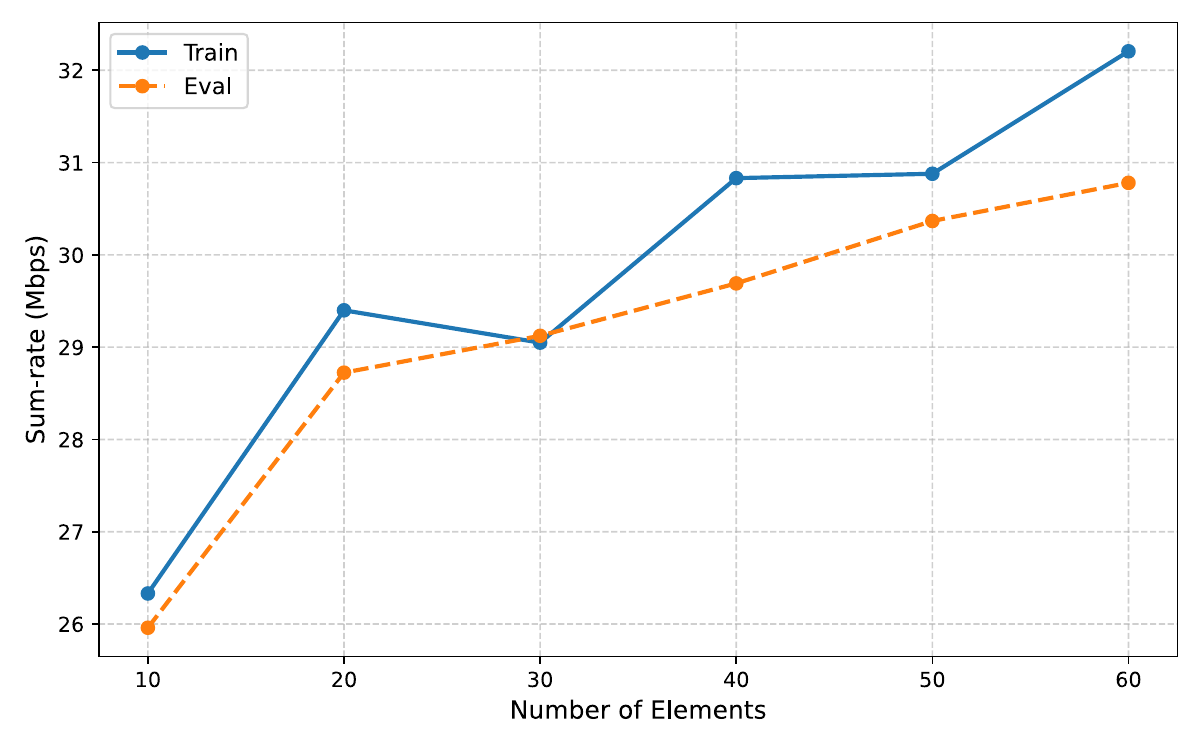}
}
\hfill
\subfloat[Harvested energy\label{fig:eh_vs_N}]{
    \includegraphics[width=0.9\linewidth]{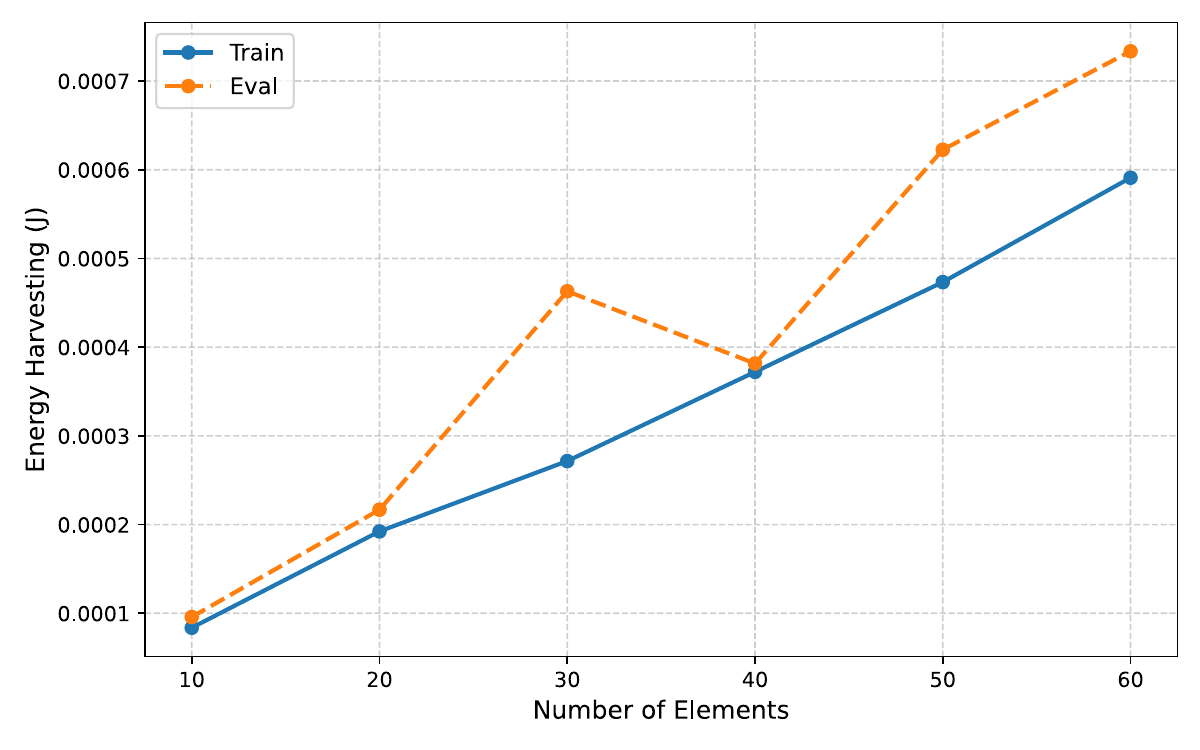}
}
\caption{Training and evaluation performance of the proposed H-PPO framework under different numbers of STAR-RIS elements $N$.}
\label{fig:combined_N}
\end{figure}

Figure~\ref{fig:combined_N} shows the influence of the number of STAR-RIS elements on the system performance for the case of the H-PPO approach. It can be seen that there is an increasing trend of both energy efficiency and sum-rate with increasing the number of elements, which is proven by Fig.~\ref{fig:combined_N}(a) and Fig.~\ref{fig:combined_N}(b), respectively. The reason behind this improvement is due to higher beamforming gains and spatial degrees of freedom with more STAR-RIS elements. The harvested energy also exhibits a monotonic increase with $N$, shown in Fig.~\ref{fig:combined_N}(c) as more reflecting elements contribute to capturing incident RF signals. This demonstrates that the system simultaneously benefits from improved communication performance and increased EH capability as the STAR-RIS size expands. Moreover, the close alignment between training and evaluation curves indicates that the learned policy generalizes well across different system scales. This confirms that H-PPO is capable of effectively exploiting the additional controllable variables introduced by larger STAR-RIS configurations.

\begin{figure}[t]
\centering
\subfloat[Energy efficiency\label{fig:ee_vs_power}]{
    \includegraphics[width=0.9\linewidth]{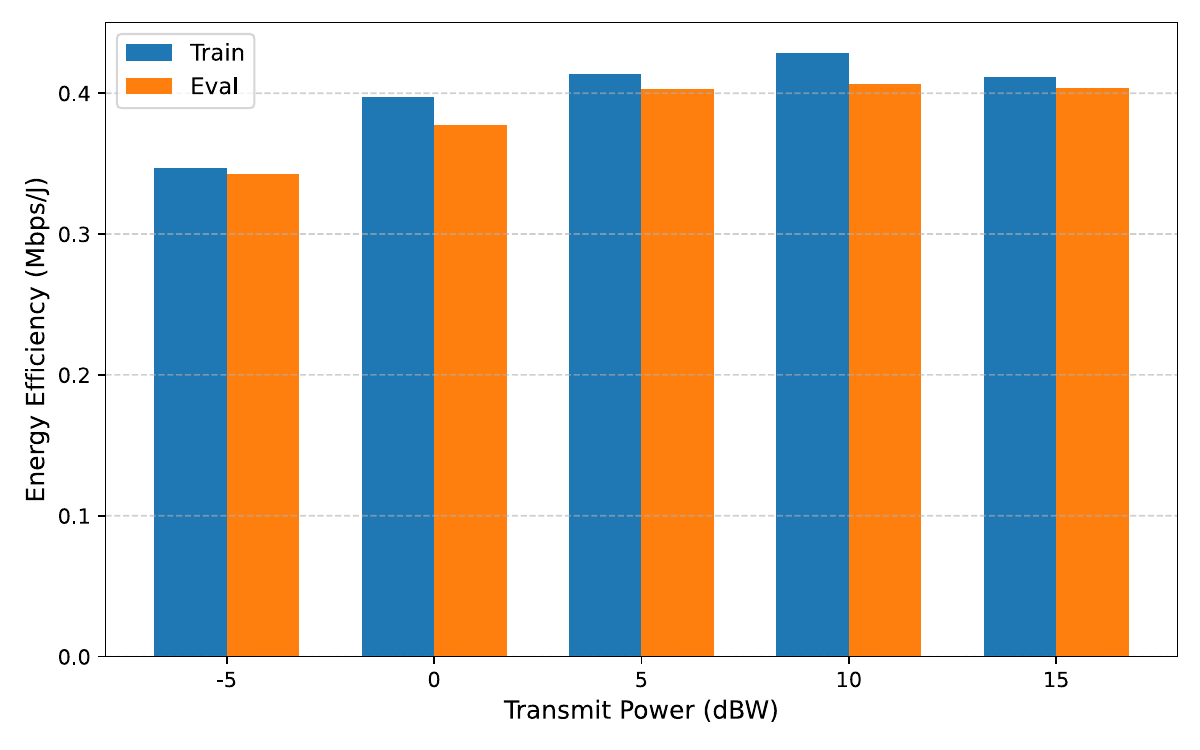}
}
\hfill
\subfloat[Sum-rate\label{fig:sumrate_vs_power}]{
    \includegraphics[width=0.9\linewidth]{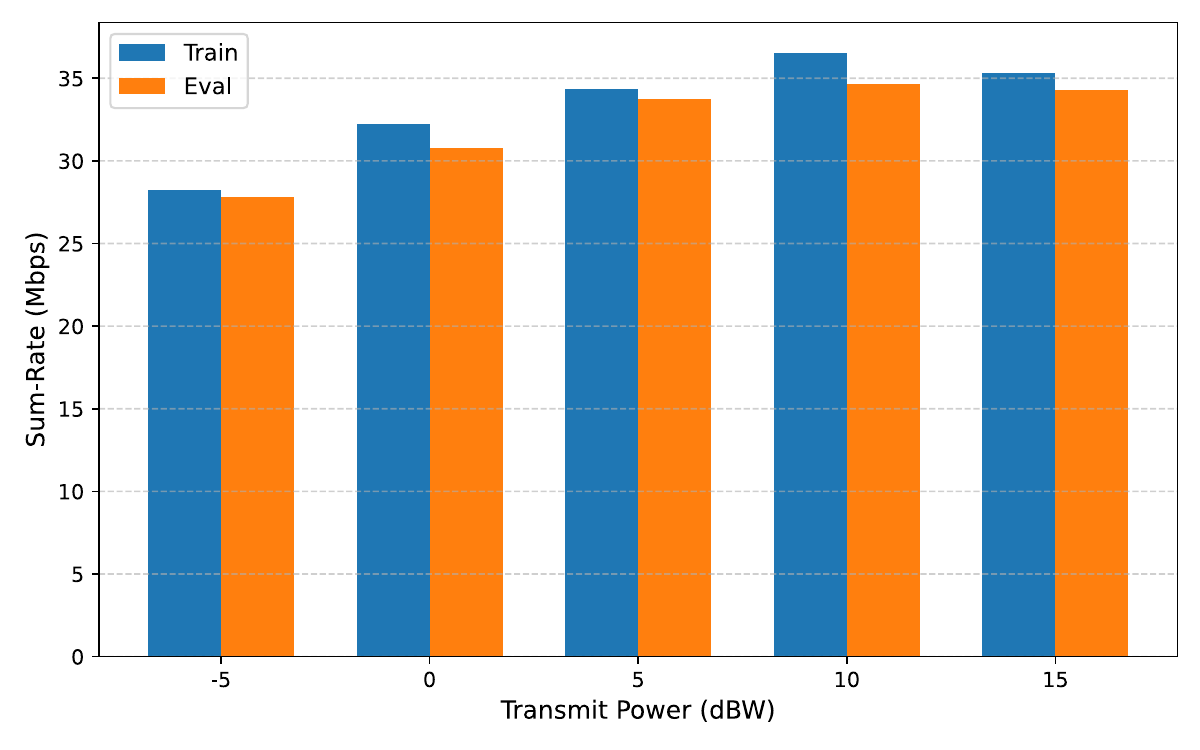}
}
\hfill
\subfloat[Harvested energy\label{fig:eh_vs_power}]{
    \includegraphics[width=0.9\linewidth]{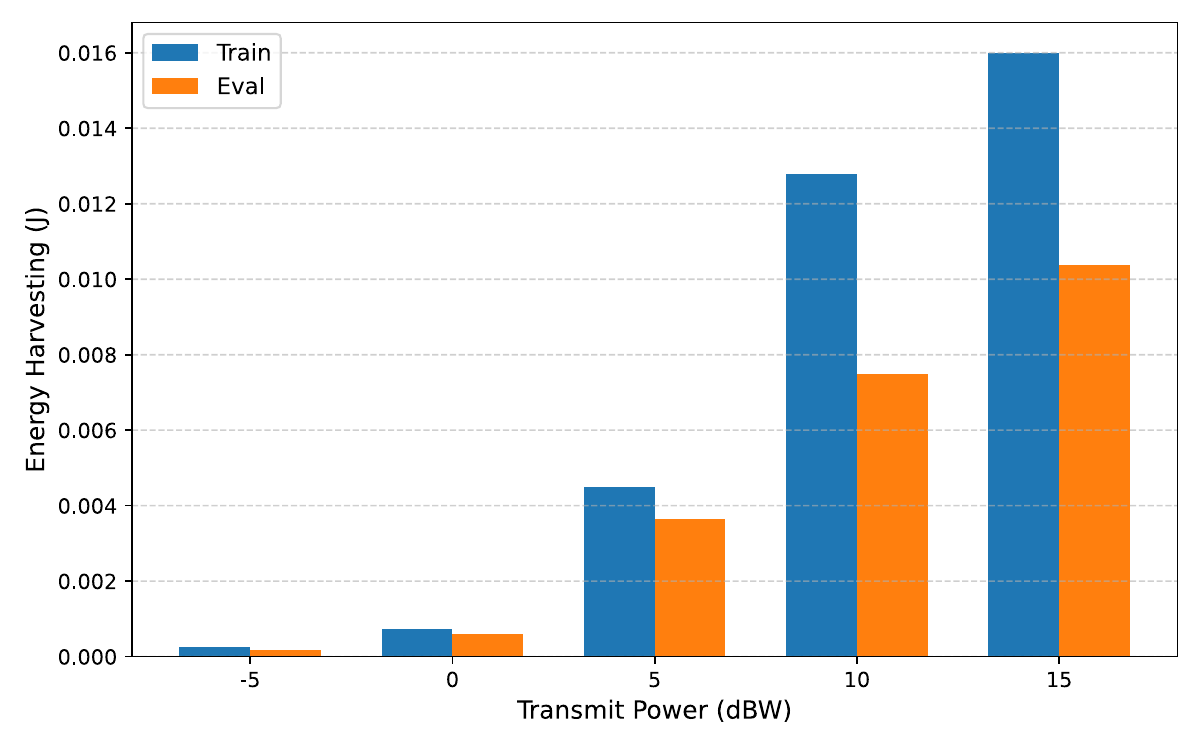}
}
\caption{Training and evaluation performance of the proposed H-PPO framework under different transmit power levels.}
\label{fig:combined_power}
\end{figure}

Figure \ref{fig:combined_power} demonstrates the impact of the transmit power on the performance of the system for the proposed H-PPO algorithm. The increase in transmit power has a positive impact on the sum rate and harvest energy due to stronger signal transmission and higher RF energy availability depicted in Fig.~\ref{fig:combined_power}(b) and Fig.~\ref{fig:combined_power}(c), respectively. Nevertheless, as shown in Fig.~\ref{fig:combined_power}(a), energy efficiency tends to saturate when there is an increase in transmit power. Although the efficiency is expected to improve with the increase in the power of transmitters at first, it starts decreasing eventually due to increased power consumption. So, it is evident that higher transmit power alone cannot yield maximum efficiency. The learned H-PPO policy successfully learns about this trend and adjusts the control variables for the best performance of the system.

\begin{figure}[t]
\centering
\subfloat[Energy efficiency\label{fig:ee_vs_users}]{
    \includegraphics[width=0.9\linewidth]{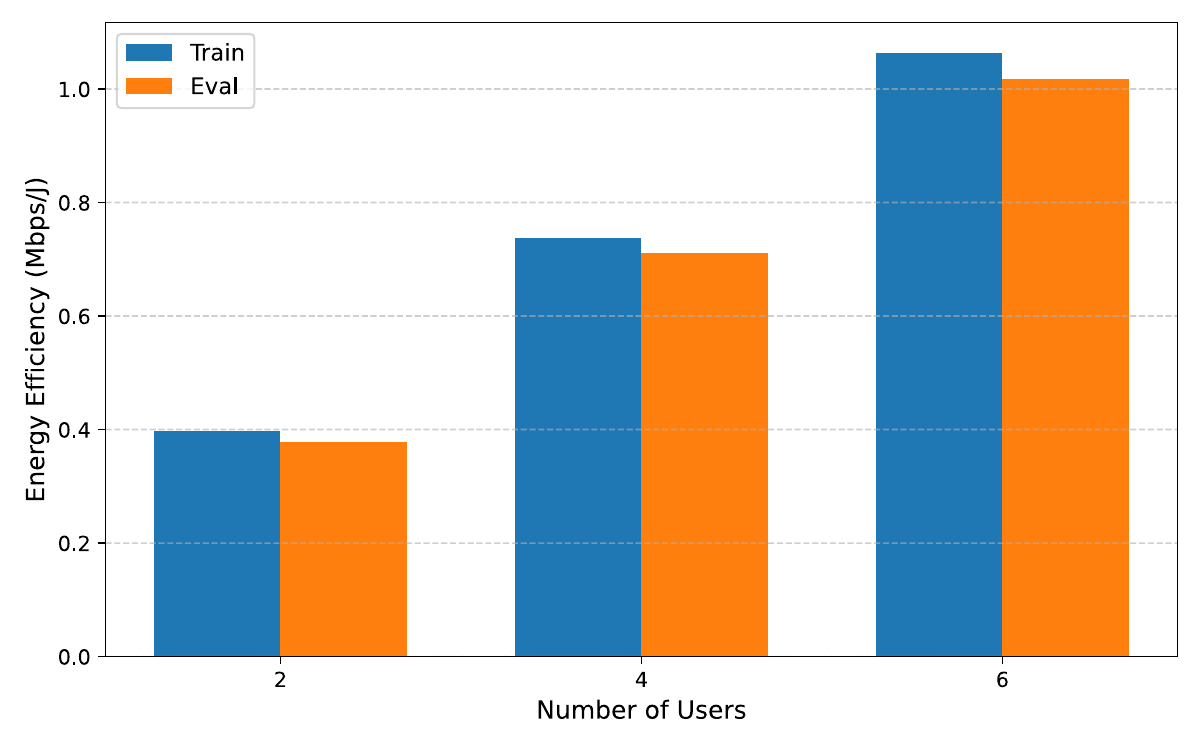}
}
\hfill
\subfloat[Sum-rate\label{fig:sumrate_vs_users}]{
    \includegraphics[width=0.9\linewidth]{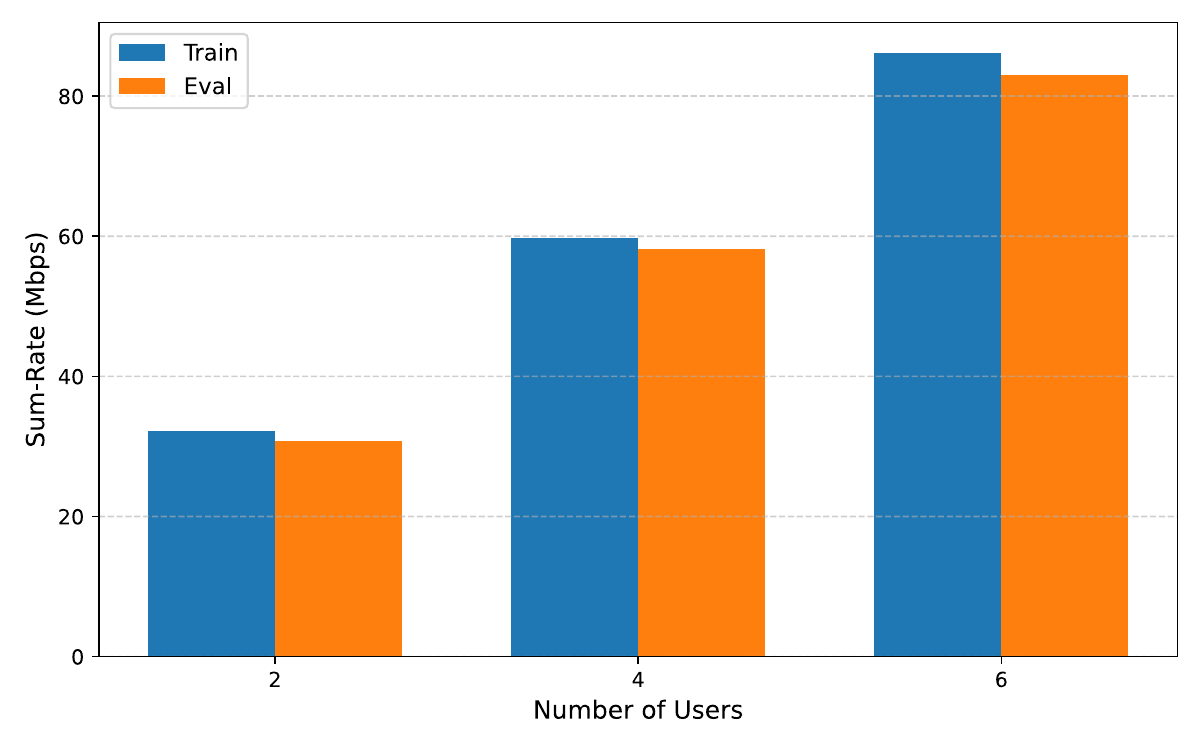}
}
\hfill
\subfloat[Harvested energy\label{fig:eh_vs_users}]{
    \includegraphics[width=0.9\linewidth]{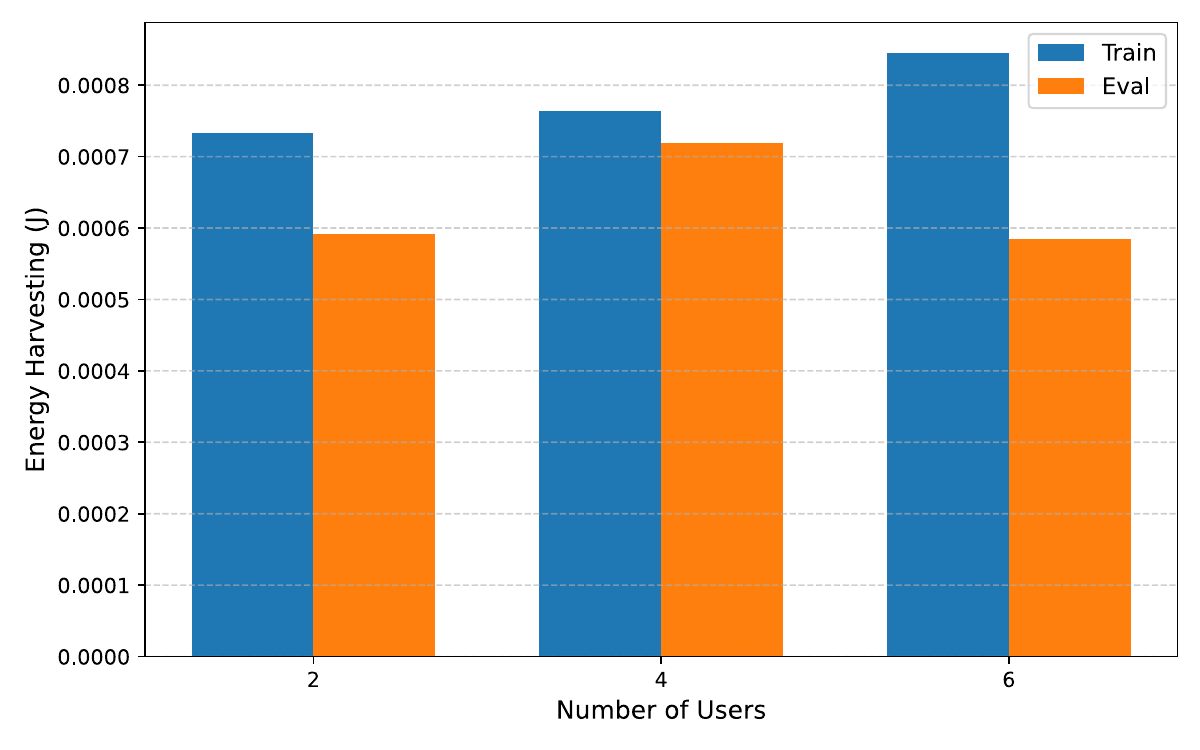}
}
\caption{Training and evaluation performance of the proposed H-PPO framework under different numbers of users.}
\label{fig:combined_users}
\end{figure}

Figure. \ref{fig:combined_users} represents the performance of the proposed algorithm versus the number of users. As observed from the simulation results, the energy efficiency and sum rate of the system tend to increase with the increase in the number of users, as shown in Fig.~\ref{fig:combined_users}(a) and Fig.~\ref{fig:combined_users}(b), respectively, thanks to the multiuser diversity gain. In addition, it can be concluded that the proposed H-PPO approach shows robust behavior even when the number of users increases. This is because the difference between training and evaluation performances is minimal. However, harvested energy is almost constant despite the increase in the number of users. This is because the learned policy is mainly aimed at maintaining communication performance and QoS satisfaction even at high user density levels.

\begin{figure*}[t]
\centering
\subfloat[Energy efficiency]{
    \includegraphics[width=0.3\linewidth]{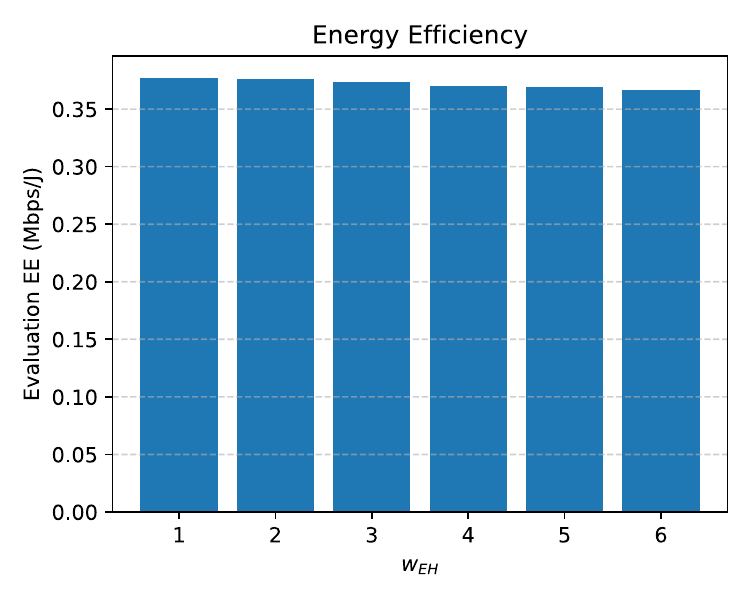}
}
\hfill
\subfloat[Sum-rate]{
    \includegraphics[width=0.3\linewidth]{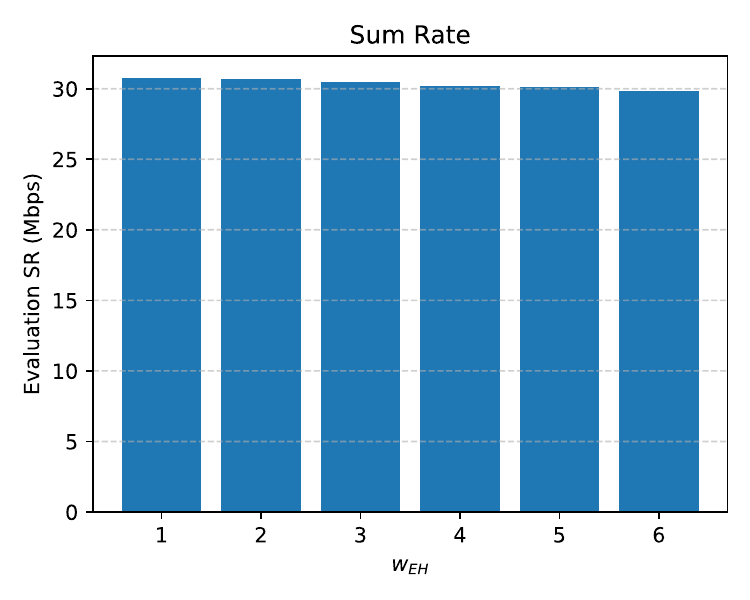}
}
\hfill
\subfloat[Harvested energy]{
    \includegraphics[width=0.3\linewidth]{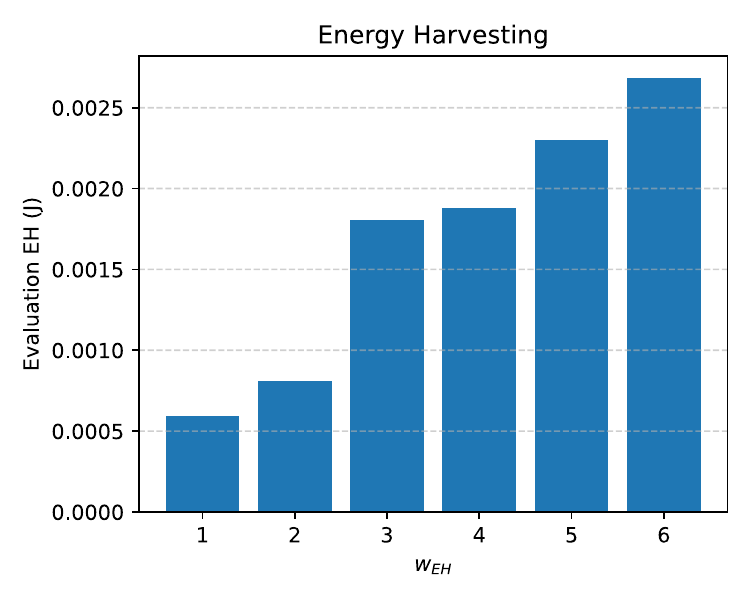}
}
\caption{Impact of the energy-harvesting reward weight $w_{\mathrm{EH}}$ on the performance of the proposed H-PPO framework with fixed $w_{\mathrm{Rate}}=1$ and $w_{\mathrm{EE}}=1$.}
\label{fig:W}
\end{figure*}

The effect of changing the reward EH weight $w_{EH}$ in the reward function on the performance of the system is illustrated in Fig.~\ref{fig:W} for $w_{Rate}$=$w_{ee}$=1. It is evident from the figure that an increase in the weight $w_{EH}$ leads to a considerable improvement in EH. The energy efficiency and sum-rate performance are almost stable due to their fixed weights. This behavior confirms that improving one objective is achieved at the expense of others in the multi-objective setting. The results demonstrate that the proposed reward formulation achieves a stable balance between communication performance and EH. When the communication-related weights (i.e., $w_{\mathrm{Rate}}$ and $w_{\mathrm{EE}}$) are fixed, the system performance remains relatively consistent, indicating that the learned policy prioritizes maintaining reliable communication performance. In contrast, varying the EH weight $w_{\mathrm{EH}}$ leads to noticeable changes in system behavior, reflecting the trade-off between communication efficiency and harvested energy.

\section{Conclusion}

This paper investigated a multi-objective optimization framework for UAV-mounted STAR-RIS-enabled wireless communication systems with integrated RF EH. A weighted optimization problem was formulated to jointly optimize the UAV trajectory, STAR-RIS transmission and reflection coefficients, BS beamforming vectors, and RF power-splitting factor under QoS, power, and mobility constraints. To solve the resulting nonconvex problem, a hybrid PPO-based DRL framework was developed by combining a categorical policy for discrete UAV movement with Beta-distributed policies for continuous communication-control variables. In addition, a two-stage reward formulation was introduced to improve training stability while balancing system sum rate, energy efficiency, and harvested energy. Simulation results validated the effectiveness of the proposed H-PPO framework, demonstrating faster convergence and more stable learning behavior than PPO and TD3 under different system configurations. The proposed framework provides an adaptive and computationally practical solution for intelligent control in UAV-mounted STAR-RIS systems. Future research directions include imperfect CSI modeling, multi-UAV cooperation, distributed DRL architectures, and hybrid RF/renewable EH integration.

\addcontentsline{toc}{section}{References}

\bibliographystyle{IEEEtran}
\bibliography{references}

@ARTICLE{9110869,
  author={Huang, Chongwen and Mo, Ronghong and Yuen, Chau},
  journal={IEEE J. Sel. Areas Commun.},
  title={{Reconfigurable intelligent surface assisted multiuser MISO systems exploiting deep reinforcement learning}},
  year={2020},
  volume={38},
  number={8},
  pages={1839--1850},
  month={Aug.},
  doi={10.1109/JSAC.2020.3000835}
}

@ARTICLE{10051712,
  author={Peng, Haoran and Wang, Li-Chun},
  journal={IEEE Trans. Wireless Commun.},
  title={{Energy harvesting reconfigurable intelligent surface for UAV based on robust deep reinforcement learning}},
  year={2023},
  volume={22},
  number={10},
  pages={6826--6838},
  month={Oct.},
  doi={10.1109/TWC.2023.3245820}
}

@ARTICLE{10458888,
  author={Aung, Pyae Sone and Nguyen, Loc X. and Tun, Yan Kyaw and Han, Zhu and Hong, Choong Seon},
  journal={IEEE Wireless Commun. Lett.},
  title={{Aerial STAR-RIS empowered MEC: A DRL approach for energy minimization}},
  year={2024},
  volume={13},
  number={5},
  pages={1409--1413},
  month={May.},
  doi={10.1109/LWC.2024.3372623}
}

@ARTICLE{9964251,
  author={Guo, Yi and Fang, Fang and Cai, Donghong and Ding, Zhiguo},
  journal={IEEE Trans. Veh. Technol.},
  title={{Energy-efficient design for a NOMA assisted STAR-RIS network with deep reinforcement learning}},
  year={2023},
  volume={72},
  number={4},
  pages={5424--5428},
  month={Apr.},
  doi={10.1109/TVT.2022.3224926}
}

@ARTICLE{10704600,
  author={Adhikari, Bhagawat and Khwaja, Ahmed Shaharyar and Jaseemuddin, Muhammad and Anpalagan, Alagan and Nallanathan, Arumugam},
  journal={IEEE Trans. Wireless Commun.},
  title={{Energy efficient RIS-assisted UAV networks using twin delayed DDPG technique}},
  year={2024},
  volume={23},
  number={12},
  pages={18423--18439},
  month={Dec.},
  doi={10.1109/TWC.2024.3468162}
}

@ARTICLE{10306287,
  author={Aung, Pyae Sone and Nguyen, Loc X. and Tun, Yan Kyaw and Han, Zhu and Hong, Choong Seon},
  journal={IEEE Internet Things J.},
  title={{Deep reinforcement learning-based joint spectrum allocation and configuration design for STAR-RIS-assisted V2X communications}},
  year={2024},
  volume={11},
  number={7},
  pages={11298--11311},
  month={Apr.},
  doi={10.1109/JIOT.2023.3329893}
}

@ARTICLE{10539342,
  author={Yao, Yuanyuan and Lv, Ke and Huang, Sai and Xiang, Wei},
  journal={IEEE Trans. Veh. Technol.},
  title={{3D deployment and energy efficiency optimization based on DRL for RIS-assisted air-to-ground communications networks}},
  year={2024},
  volume={73},
  number={10},
  pages={14988--15003},
  month={Oct.},
  doi={10.1109/TVT.2024.3405608}
}

@article{CHAUDHARY2025700,
  author={Chaudhary, Shivam and Budhiraja, Ishan and Chaudhary, Rajat and Garg, Sahil and Choi, Bong Jun and Alrashoud, Mubarak},
  journal={Alex. Eng. J.},
  title={{Proximal policy optimization-based sum rate maximization scheme for STAR-RIS-assisted vehicular networks underlaying UAV}},
  year={2025},
  volume={118},
  pages={700--710},
  month={Apr.},
  doi={10.1016/j.aej.2024.12.115}
}

@article{KHALIL2023109784,
  author={Khalil, Alvi Ataur and Selim, Mohamed Y. and Rahman, Mohammad Ashiqur},
  journal={Comput. Netw.},
  title={{Deep learning-based energy harvesting with intelligent deployment of RIS-assisted UAV-CFmMIMOs}},
  year={2023},
  volume={229},
  pages={109784},
  month={Jun.},
  doi={10.1016/j.comnet.2023.109784}
}

@ARTICLE{9741332,
  author={Wang, Tianqi and Fang, Fang and Ding, Zhiguo},
  journal={IEEE Trans. Veh. Technol.},
  title={{An SCA and relaxation based energy efficiency optimization for multi-user RIS-assisted NOMA networks}},
  year={2022},
  volume={71},
  number={6},
  pages={6843--6847},
  month={Jun.},
  doi={10.1109/TVT.2022.3162197}
}

@ARTICLE{10187161,
  author={Yuan, Xin and Hu, Shuyan and Ni, Wei and Wang, Xin and Jamalipour, Abbas},
  journal={IEEE Trans. Inf. Forensics Security},
  title={{Deep reinforcement learning-driven reconfigurable intelligent surface-assisted radio surveillance with a fixed-wing UAV}},
  year={2023},
  volume={18},
  pages={4546--4560},
  month={Jul.},
  doi={10.1109/TIFS.2023.3297021}
}

@ARTICLE{10187159,
  author={Gao, Xinyu and Yi, Wenqiang and Liu, Yuanwei and Zhang, Jianhua and Zhang, Ping},
  journal={IEEE Trans. Commun.},
  title={{DRL-enabled coverage and capacity optimization in STAR-RIS-assisted networks}},
  year={2023},
  volume={71},
  number={11},
  pages={6616--6632},
  month={Nov.},
  doi={10.1109/TCOMM.2023.3296753}
}

@ARTICLE{10802901,
  author={Li, Jiameng and Huang, Yihang and Wu, Jun and Wang, Xiaojie and Ning, Zhaolong},
  journal={IEEE Internet Things J.},
  title={{Energy-efficiency maximization for STAR-RIS and AAV-assisted IUA: A multiagent DRL approach}},
  year={2025},
  volume={12},
  number={21},
  pages={43936--43948},
  month={Nov.},
  doi={10.1109/JIOT.2024.3518081}
}

@ARTICLE{10793392,
  author={Lin, Na and Wu, Tianxiong and Hawbani, Ammar and Zhao, Liang and Liu, Chunxiao and Wan, Shaohua and Guizani, Mohsen},
  journal={IEEE Internet Things J.},
  title={{Energy-efficiency optimization in RIS-assisted AAV communications based on deep reinforcement learning}},
  year={2025},
  volume={12},
  number={8},
  pages={11036--11048},
  month={Apr.},
  doi={10.1109/JIOT.2024.3515108}
}

@ARTICLE{10452297,
  author={Yang, Helin and Lin, Kailong and Xiao, Liang and Zhao, Yifeng and Xiong, Zehui and Han, Zhu},
  journal={IEEE Trans. Wireless Commun.},
  title={{Energy harvesting UAV-RIS-assisted maritime communications based on deep reinforcement learning against jamming}},
  year={2024},
  volume={23},
  number={8},
  pages={9854--9868},
  month={Aug.},
  doi={10.1109/TWC.2024.3367034}
}

@ARTICLE{9837935,
  author={Zhong, Ruikang and Liu, Yuanwei and Mu, Xidong and Chen, Yue and Wang, Xianbin and Hanzo, Lajos},
  journal={IEEE J. Sel. Areas Commun.},
  title={{Hybrid reinforcement learning for STAR-RISs: A coupled phase-shift model based beamformer}},
  year={2022},
  volume={40},
  number={9},
  pages={2556--2569},
  month={Sep.},
  doi={10.1109/JSAC.2022.3192053}
}

@ARTICLE{9645164,
  author={Nguyen, Khoi Khac and Khosravirad, Saeed R. and da Costa, Daniel Benevides and Nguyen, Long D. and Duong, Trung Q.},
  journal={IEEE J. Sel. Topics Signal Process.},
  title={{Reconfigurable intelligent surface-assisted multi-UAV networks: Efficient resource allocation with deep reinforcement learning}},
  year={2022},
  volume={16},
  number={3},
  pages={358--368},
  month={Apr.},
  doi={10.1109/JSTSP.2021.3134162}
}

@ARTICLE{11174947,
  author={Salim, Mahmoud M. and Rabie, Khaled M. and Muqaibel, Ali H.},
  journal={IEEE Commun. Lett.},
  title={{Robust energy-efficient DRL-based optimization in UAV-mounted RIS systems with jitter}},
  year={2025},
  volume={29},
  number={12},
  pages={2780--2784},
  month={Dec.},
  doi={10.1109/LCOMM.2025.3612448}
}

@ARTICLE{11215794,
  author={Salim, Mahmoud M. and Rabie, Khaled M. and Muqaibel, Ali H.},
  journal={IEEE Commun. Stand. Mag.},
  title={{Energy-efficient UAV-mounted RIS for IoT: A hybrid energy harvesting and DRL approach}},
  year={2025},
  pages={1--9},
  month={Oct.},
  doi={10.1109/MCOMSTD.2025.3620005}
}

@ARTICLE{10439000,
  author={Wang, Siqiang and Zheng, Zhong and Fei, Zesong and Guo, Jing and Yuan, Jinhong and Sun, Zhi},
  journal={IEEE Trans. Wireless Commun.},
  title={{Toward ergodic sum rate maximization of multiple-RIS-assisted MIMO multiple access channels over generic Rician fading}},
  year={2024},
  volume={23},
  number={8},
  pages={9613--9628},
  month={Aug.},
  doi={10.1109/TWC.2024.3364158}
}

@ARTICLE{11176824,
  author={Gan, Wei and Peng, Yuyang and Ye, Runlong and Al-Hazemi, Fawaz and Lee, Juho},
  journal={IEEE Trans. Cogn. Commun. Netw.},
  title={{STAR-RIS aided SWIPT systems with non-linear energy harvesting model}},
  year={2026},
  volume={12},
  pages={1327--1341},
  month={Sep.},
  doi={10.1109/TCCN.2025.3613576}
}

@ARTICLE{10891707,
  author={Yaswanth, Jetti and Saikia, Prajwalita and Singh, Keshav and Kim, Yun Hee and Duong, Trung Q.},
  journal={IEEE Trans. Cogn. Commun. Netw.},
  title={{Active and STAR-RIS-assisted MIMO ISAC systems with SWIPT}},
  year={2025},
  volume={11},
  number={6},
  pages={3842--3858},
  month={Dec.},
  doi={10.1109/TCCN.2025.3543405}
}

@ARTICLE{11214186,
  author={Ata, Yalçın and Leoni, Claudia and Vegni, Anna Maria},
  journal={IEEE Trans. Commun.},
  title={{Time switching-variable mode splitting protocol for STAR-RIS-aided SWIPT in NOMA networks}},
  year={2026},
  volume={74},
  pages={247--260},
  month={Oct.},
  doi={10.1109/TCOMM.2025.3624160}
}

@article{schulman2017ppo,
  author={Schulman, John and Wolski, Filip and Dhariwal, Prafulla and Radford, Alec and Klimov, Oleg},
  journal={arXiv preprint arXiv:1707.06347},
  title={{Proximal policy optimization algorithms}},
  year={2017}
}

@article{WU2025103744,
  author={Wu, Wenjie and Luo, Zhongqiang},
  journal={Ain Shams Eng. J.},
  title={{Computation offloading in STAR-RIS aided U-MEC systems: A delay-minimization deep reinforcement learning approach}},
  year={2025},
  volume={16},
  number={12},
  pages={103744},
  month={Dec.},
  doi={10.1016/j.asej.2025.103744}
}

@ARTICLE{8663615,
  author={Zeng, Y. and Xu, J. and Zhang, R.},
  journal={IEEE Trans. Wireless Commun.},
  title={{Energy minimization for wireless communication with rotary-wing UAV}},
  year={2019},
  volume={18},
  number={4},
  pages={2329--2345},
  month={Apr.},
  doi={10.1109/TWC.2019.2902559}
}

@ARTICLE{10361836,
  author={Zhou, Hao and Erol-Kantarci, Melike and Liu, Yuanwei and Poor, H. Vincent},
  journal={IEEE Commun. Surv. Tutor.},
  title={{A survey on model-based, heuristic, and machine learning optimization approaches in RIS-aided wireless networks}},
  year={2024},
  volume={26},
  number={2},
  pages={781--823},
  month={Jun.},
  doi={10.1109/COMST.2023.3340099}
}

@ARTICLE{11207621,
  author={Salim, M. M. and Rabie, K. M. and Muqaibel, A. H.},
  journal={IEEE Internet Things J.},
  title={{Energy-efficient irregular RIS-aided UAV-assisted optimization: A deep reinforcement learning approach}},
  year={2025},
  volume={12},
  number={24},
  pages={55196--55210},
  month={Dec.},
  doi={10.1109/JIOT.2025.3623306}
}

\end{document}